\documentclass[bibyear]{aa}

\usepackage{natbib}
\usepackage{hyperref}
\usepackage{xcolor}
\usepackage{orcidlink}
\usepackage{graphicx}
\usepackage{txfonts}
\usepackage{booktabs}
\usepackage{lipsum}
\usepackage{comment}

\newcommand{\athenapk}{{\sc AthenaPK}}

\newcommand{\bondi}{\texttt{bondi}}
\newcommand{\cool}{\texttt{cool}}
\newcommand{\turbolow}{\texttt{turb\_low}}
\newcommand{\turbohigh}{\texttt{turb\_high}}
\newcommand{\lowc}{\texttt{cca\_low}}
\newcommand{\highc}{\texttt{cca\_high}}

\hypersetup{
    colorlinks=true,
    linkcolor=blue,
    citecolor=blue,
    filecolor=blue,
    urlcolor=blue,
    menucolor=blue
}

\begin{document}

\title{{\bfseries\scshape BlackHoleWeather –} Chaotic cold accretion across the meso-scale: 
{\Large Variability and kinematics}}

\titlerunning{Variability and kinematics of multiphase accretion}

   \authorrunning{Filippo Barbani et al.}
     
   \author{Filippo Barbani \thanks{filippo.barbani@unimore.it}
        \inst{1}\orcidlink{0000-0002-1620-2577},
        Massimo Gaspari
        \inst{1}\orcidlink{0000-0003-2754-9258},
        Olmo Piana
        \inst{1}\orcidlink{0000-0002-1558-5289},
        Vieri Cammelli
        \inst{1}\orcidlink{0000-0002-2070-9047},
        Fred J. Jennings
        \inst{1}\orcidlink{0009-0000-0152-9983},      
        Davide M. Brustio
        \inst{1}\orcidlink{0009-0009-7700-1910},
        Giovanni Stel
        \inst{1}\orcidlink{0009-0007-0585-9462},          
        Valeria Olivares
         \inst{2, 3}\orcidlink{0000-0001-6638-4324},
        Filippo M. Maccagni
         \inst{4, 5}\orcidlink{0000-0002-9930-1844},
        Martin Fournier
         \inst{6}\orcidlink{0009-0006-2593-1583}, 
        Francesco Tombesi
         \inst{7, 8, 9}\orcidlink{0000-0002-6562-8654},        
        Pasquale Temi
         \inst{10}\orcidlink{0000-0002-8341-342X},
        Fabrizio Fiore
         \inst{11}\orcidlink{0000-0002-4031-4157},  
        Roberto Serafinelli
         \inst{7, 12}\orcidlink{0000-0001-6638-4324}, 
        \and
        Ashkbiz Danehkar
         \inst{13}\orcidlink{0000-0003-4552-5997}
          }

   \institute{
    Department of Physics, Informatics and Mathematics, University of Modena and Reggio Emilia, I-41125 Modena, Italy
   \and
   Department of Physics, Universidad de Santiago de Chile, Santiago, Chile
    \and
    CIRAS, Universidad de Santiago de Chile, Santiago, Chile
    \and
    INAF -- Osservatorio Astronomico di Cagliari, via della Scienza 5, 09047, Selargius (CA), Italy
    \and
    Wits Centre for Astrophysics, School of Physics, University of the Witwatersrand, 2000, Johannesburg, South Africa
    \and
    Universit\"{a}t Hamburg, Hamburger Sternwarte, Gojenbergsweg 112, 21029 Hamburg, Germany
    \and
    INAF -- Astronomical Observatory of Rome, 00078 Monte Porzio Catone (Rome), Italy
    \and
    Department of Physics, University of Rome ``Tor Vergata'', 00133 Rome, Italy
    \and
    INFN -- Rome ``Tor Vergata'' Section, 00133 Rome, Italy
    \and
    NASA Ames Research Center, MS 245-6, Moffett Field, CA 94035-1000, USA
    \and
    INAF -- Astronomical Observatory of Trieste, 34143 Trieste, Italy
    \and
    Instituto de Estudios Astrof\'isicos, Facultad de Ingenier\'ia y Ciencias, Universidad Diego Portales, Avenida Ej\'ercito Libertador 441, Santiago, Chile
    \and
    Science and Technology Institute, Universities Space Research Association, Huntsville, AL 35805, USA
   }
   
   \date{\today}


 
\abstract
{Accretion onto supermassive black holes (SMBHs) in realistic halos is inherently time-variable, governed by the interplay of turbulence, radiative cooling, and multiphase condensation. In chaotic cold accretion (CCA), multiphase clouds and filaments condense out of the hot atmosphere and feed the central SMBH stochastically.}
{We investigate how varying turbulence regulates the variability, radial transport, and kinematics of CCA, focusing on the meso-scale that connects halo rain to the inner inflow.}
{We analyse 3D hydrodynamic simulations with a GPU-accelerated code, including radiative cooling and driven subsonic turbulence in a stratified galaxy group, resolving scales from tens of kpc down to sub-pc and probing two turbulent weather regimes.}
{In both regimes, SMBH accretion proceeds through CCA, remains strongly super-Bondi, and varies by up to $\sim$\,2 dex. The runs diverge mainly at meso-scales: strong stirring sustains an extended, fragmented multiphase feeding and a clear inflow enhancement at 0.1--1 kpc, whereas weaker turbulence yields a smoother, more centrally concentrated cascade. Yet the innermost feeding rates remain similar, implying that SMBH accretion is not directly supply-limited by the macro-scale weather. The accretion rate distributions peak at low Eddington ratios, indicating a predominantly maintenance-mode state. The accretion rate power spectra follow a broken power law, with pink noise on long/medium timescales and a steeper red-noise tail at high frequencies, consistent with a parsec-scale collisional damping transition. CCA modes are well captured by two complementary diagnostics: the $\mathcal{C}$-ratio ($\equiv t_{\rm cool}/t_{\rm eddy}$) $\approx 1$ identifies the soft X-ray gas as the gateway of the condensation cascade, while the k-plot (line broadening vs.~shift) shows that the weather distinction is strongest on meso-scales, where the stormy regime produces broader, more overlapping multiphase kinematics than the rainy regime.}
{The meso-scale is the critical bridge between halo rain and micro-scale CCA feeding, regulating the spatial transport, kinematic imprint, and temporal coherence of the SMBH growth.}


   \maketitle
%
\section{Introduction}

Most massive galaxies host a central supermassive black hole (SMBH), whose growth is linked to the evolution of its host galaxy and surrounding halo \citep{Kormendy2013}. Through accretion-powered activity, SMBHs release large amounts of energy and momentum in the form of radiation, winds, and relativistic jets, giving rise to the so-called active galactic nuclei \citep[AGN,][]{Lynden-Bell1969}. This feedback plays a fundamental role in regulating gas cooling, star formation, and the thermodynamic structure of galactic halos, thereby shaping the observed galaxy population across cosmic time \citep[e.g.][]{mcnamara2012,GittiBrighenti2012,king2015,fiore2017}.

On large scales, most baryons reside in diffuse, hot gaseous atmospheres—the circumgalactic, intragroup, and intracluster medium \citep[CGM, IGrM, ICM,][]{WhiteFrenk1991,kravtsov2012,tumlinson2017}, which behave as stratified plasmas governed by the interplay between gravity, turbulence, radiative cooling, and feedback. Far from being smooth and static, these halos exhibit complex thermodynamic and kinematic structure. Multi-wavelength observations increasingly reveal that hot halos host multiphase gas, including warm ionised and cold atomic and molecular components, frequently arranged in extended filamentary structures \citep[e.g.][]{fabian2008,McDonaldVeilleux2011,tremblay2016,tremblay2018,maccagni2021, morganti2023, ubertosi2023,ubertosi2025,toni2026}. These filaments can extend tens of kiloparsecs from the galaxy centre and are often spatially correlated with AGN jets and cavities, indicating a close coupling between cooling gas and feedback activity. The formation and evolution of multiphase gas is one of the key open questions
in extragalactic astrophysics \citep[e.g.][]{McNamaraRussell2016,VoitMeece2017, waters2019, GaspariTombesi2020}. Indeed, large-scale condensation can launch a precipitation cascade (akin to Earth weather) of cold clouds and filaments \citep[e.g.][]{wittor2020,wittor2023} that delivers multiphase inflows towards the nucleus. A key open step is to connect this halo-scale condensation cascade to the time-domain statistics and kinematic signatures of SMBH fueling across the meso-scale (parsecs to kiloparsecs) bridge.

In a stratified atmosphere, turbulence seeds density fluctuations; when cooling becomes competitive with turbulent stirring, the densest perturbations survive mixing, decouple from the hot phase, and condense into warm and cold structures via nonlinear thermal instability.
This behaviour is commonly described within the chaotic cold accretion (CCA) framework, in which turbulence and thermal instability promote the condensation of cold clouds and filaments that rain stochastically onto the SMBH \citep{gaspari2013,gaspari2015,gaspari2017,gaspari2017b,Barbani2026a}. In this picture, SMBH fueling is intrinsically clumpy, anisotropic, and time-dependent. Collisions between cold clouds and filaments efficiently cancel angular momentum, enhancing accretion rates relative to classical Bondi inflow and naturally producing strong variability. CCA provides a coherent physical framework linking halo thermodynamics, multiphase structure, and AGN fueling, and is supported by growing observational evidence across radio, millimeter, optical, ultraviolet, and X-ray wavelengths \citep[e.g.][]{MaccagniMorganti2014,voit2015,Mcdonald2018,tremblay2018,maccagni2021,temi2022,olivaressalome2022,olivares2025,reefe2025,romero2025,omoruyi2026}.
This CCA picture is distinct from, but complementary to, sub-pc chaotic-accretion models, where SMBHs grow through randomly oriented accretion episodes linked to galaxy interactions \citep{KingPringle2006,KingPringle2008}.

The intrinsically stochastic nature of multiphase accretion naturally links to AGN variability, reflecting the time-dependent and clumpy character of SMBH fueling. Variability is observed across the electromagnetic spectrum and over a wide range of timescales, from hours to decades and beyond \citep[see][for a review]{PaolilloPapadakis2025}. Typical luminosity fluctuations reach amplitudes of $10$–$30\%$ on timescales of days to months, increasing to factors of a few on yearly to decadal scales \citep[e.g.][]{McHardyKoerding2006,KellyBechtold2009,MushotzkyEdelson2011, CammelliTan2025}. In extreme cases, such as changing-look AGN, optical and X-ray fluxes can vary by one to two orders of magnitude, implying dramatic changes in the instantaneous accretion rate onto the SMBH \citep{lamassa2015,runnoe2016,macleod2019,TozziLusso2022}. Recent studies indicate that these variations are predominantly driven by intrinsic changes in $\dot{M}_\bullet$, rather than by variable obscuration along the line of sight \citep[e.g.][]{yang2023,lyu2025}. On longer timescales, AGN activity is also highly intermittent, as revealed by X-ray cavities, radio lobes, and ionisation echoes that trace duty cycles of $10^4$–$10^6$ yr \citep{keel2012,Schawinski2015,french2023}. Such flickering behaviour is also consistent with theoretical expectations, which predicts characteristic feeding episodes of $\sim 10^5$ yr \citep{KingNixon2015}. The amplitude of variability generally increases with timescale, consistent with the idea that longer-term fluctuations originate from processes operating on progressively larger spatial scales and involving more massive gas reservoirs \citep{ArevaloUttley2006,VagnettiMiddei2016}. These observations suggest that AGN variability encodes information about the physical mechanisms governing gas inflow across a broad range of radii, from the halo to the innermost accretion flow.

Numerical hydrodynamical simulations have become a cornerstone of modern astrophysics \citep[see e.g.][for a review]{Vogelsberger2020,bourne2023}, enabling the study of nonlinear, multi-scale processes across a wide range of systems and physical regimes, such as star-forming galaxies \citep[e.g.][]{hopkins2018,pillepich2019,hopkins2023,barbani2023,barbani2025} and more massive quiescent galaxies, groups and clusters \citep[e.g.][]{gaspari2012,fournier2025,sotira2026}. In the AGN context, high-resolution simulations are essential to capture the interplay between cooling, turbulence, and feedback that drives multiphase condensation and stochastic SMBH fueling. Significant progress has also been made in recent multiscale simulations of AGN accretion and feedback \citep[e.g.][]{Beckmann2019,anglesalcaraz2021,bourne2021,guo2025,shin2025}. Despite this progress, fully resolving the cascade from halo scales down to sub-parsec accretion remains computationally and physically challenging. Most numerical simulations resolve inflows only down to parsec or tens-of-parsec scales and have primarily focused on massive cluster environments, limiting direct predictions for accretion variability.

This study is conducted within the {\sc BlackHoleWeather} project (PI: Gaspari), which targets a unified, multi-physics description of SMBH feeding and AGN self-regulation across environments ranging from galaxies to groups and clusters. {\sc BlackHoleWeather} integrates high-resolution simulations with synthetic multi-wavelength observables and theory-driven diagnostics to connect multiphase halo weather, accretion variability, and feedback in a single consistent framework.
Building on the controlled stratified-halo experiments that underpin the CCA picture \citep{gaspari2013,gaspari2017}, we examine turbulence-induced condensation and SMBH fueling down to sub-pc scales in a hot, stratified atmosphere representative of group-scale haloes. The companion paper \citep[][B26a hereafter]{Barbani2026a} presents the multiphase morphology and thermodynamics across the transitional meso-scale (pc to kpc) linking halo precipitation to the inner inflow. Here we concentrate on variability and kinematics, quantifying the temporal statistics and power spectra of the SMBH accretion rate and connecting them to CCA diagnostics (e.g. k-plots and $\mathcal{C}$-ratios; \citealt{gaspari2018}) that trace the competition between condensation and turbulence.

We use the same numerical setup analysed in B26a. We employ \athenapk, a GPU-boosted adaptive-mesh-refinement (AMR) hydrodynamics code, to follow the halo atmosphere from large scales to sub-parsec radii. By varying the turbulence driving strength, we explore how different halo-weather regimes regulate the time variability of SMBH inflow, affecting intermittency and characteristic timescales, and how these temporal signatures are connected to the emergent multiphase structure.
Within {\sc BlackHoleWeather}, the project activities are grouped into thematic work packages spanning multiscale feeding and feedback (cf.~Figure 1 in \citealt{GaspariTombesi2020}). The present study contributes to the feeding-focused package (WP2), centred on turbulence-driven condensation and multiscale accretion onto the SMBH. 

This paper is structured as follows. In Section~\ref{num} we summarise the numerical setup and physical prescriptions. In Section~\ref{sec:accretion} we analyse the SMBH accretion histories and compare the two turbulence regimes. Section~\ref{sec:timevariability} discusses the statistical properties of the variability and the corresponding power spectra. Section~\ref{cca_diag} introduces CCA diagnostics linking the competition between cooling and turbulence to the halo thermodynamic state. Section~\ref{sec:synthesis} compares the results with previous studies, and Section~\ref{conc} summarises the conclusions and future perspectives.

\begin{figure*}
\centering
\includegraphics[width=1\textwidth]{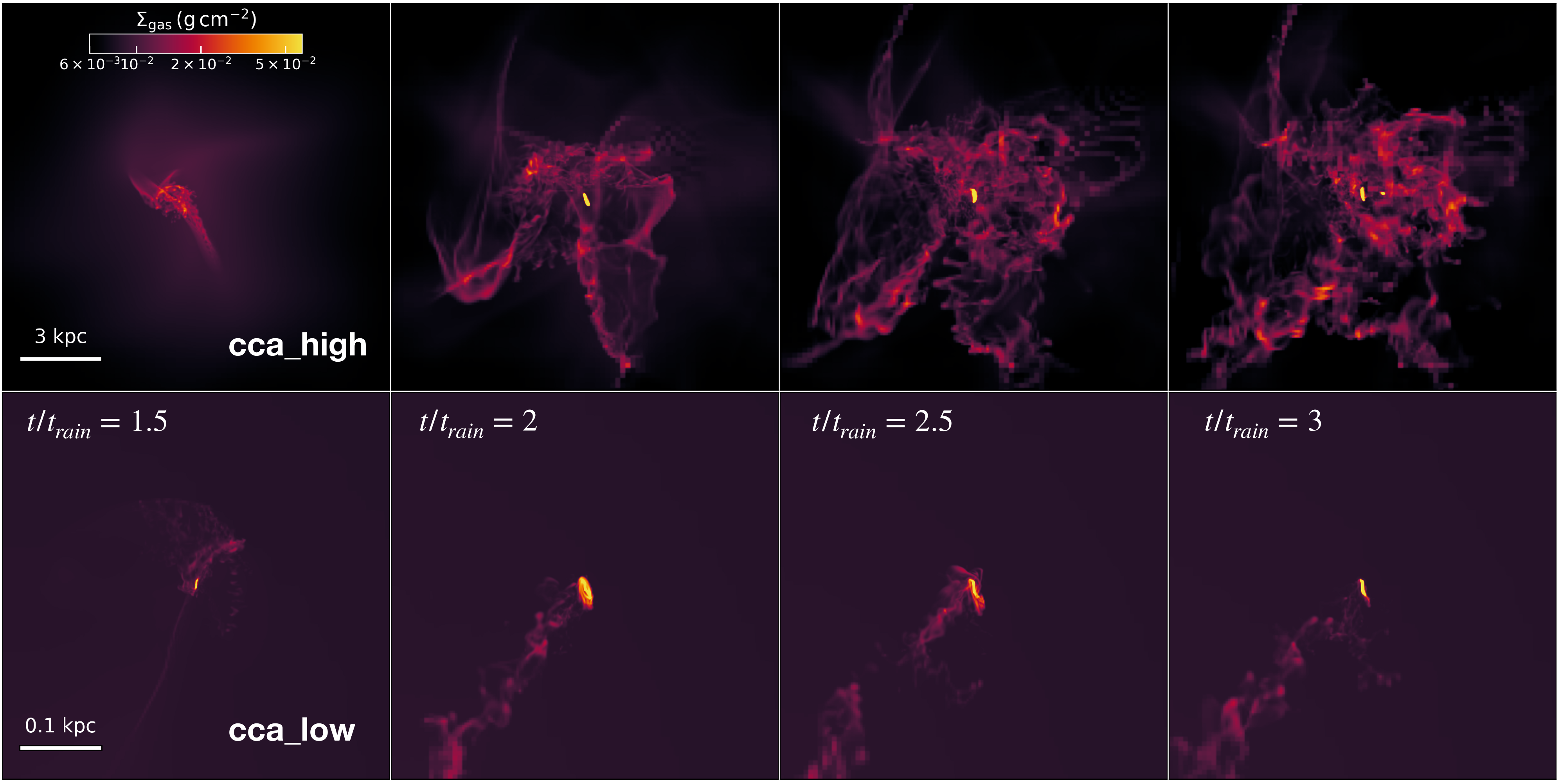}

\caption{Evolution of the projected gas surface density
$\Sigma_{\rm gas}$ in the \highc\ (top panels) and
\lowc\ (bottom panels) simulations at four times normalised by $t_{\rm rain}$. Projections are computed along the $x$-axis. The top row spans a 15~kpc region, highlighting the extended multiphase structure, while the bottom row focuses on the central 0.5~kpc, emphasising the inner condensation and inflow.}\label{tempslice}

\end{figure*}

\section{Physics and numerical methods}\label{num}
The implemented physics and numerics are described in detail in B26a, to which we refer the reader for a complete explanation of the processes used in the simulations. Here we provide a concise overview of the numerical setup and physical models.

All simulations are performed with the AMR code \athenapk, a performance-portable magnetohydrodynamics code based on the Athena++/Parthenon/Kokkos framework \citep{stone2020,grete23,edwards2014,trott21}. The code solves the Euler equations of hydrodynamics in conservative Eulerian form using a second-order Godunov scheme \citep{godunov1959} with piecewise-linear reconstruction \citep{vanLeer1979} and an HLLC Riemann solver \citep{toro1994}, with time integration performed via a second-order Runge--Kutta scheme \citep{butcher2008}. To ensure numerical robustness in the presence of strong cooling and multiphase flows, a local first-order fallback scheme is applied whenever non-physical states arise. 
Magnetic fields are not included in the present simulations, in order to isolate the hydrodynamic effect of turbulent stirring on the CCA cascade. This is a controlled hydrodynamical limit: the hot IGrM/ICM is expected to be globally high-$\beta$, with magnetic pressure sub-dominant to thermal pressure, although magnetic fields may still affect the morphology, survival, and kinematics of the locally condensed phase. Their impact will be explored in upcoming {\sc BlackHoleWeather} work.

The computational domain is a cubic box of size $L_{\rm box}=50$ kpc, discretized using a static mesh refinement (SMR) hierarchy of nested Cartesian grids centred on the origin. The root grid contains $128^3$ cells, and refinement proceeds through 12 fixed levels, each increasing the spatial resolution by a factor of two towards the centre. This setup yields a finest spatial resolution of $\Delta x_{\min}\simeq0.1$ pc within the inner $\sim10$ pc, while efficiently capturing large-scale IGrM dynamics on coarser grids. The SMR progressively increases the resolution towards smaller radii, such that the cooling length (i.e. $\ell_{\rm cool}\sim c_s\,t_{\rm cool}$, where $c_s$ is the sound speed and $t_{\rm cool}$ is the cooling time) of the cooling gas approaching condensation is well resolved in the central $\sim 10$ kpc region\footnote{In the presence of turbulence, the classical cooling length should be interpreted as a lower estimate, since turbulent transport can broaden the effective scale over which thermally unstable gas cools and condenses.}. The finest fragmentation of the cold phase, however, may still retain some resolution dependence. Each grid block contains $32^3$ cells, and standard prolongation and restriction operators ensure consistency across refinement boundaries.

The simulations are initialized with a hot gaseous halo in hydrostatic equilibrium embedded in a static gravitational potential. The potential includes contributions from a Navarro--Frenk--White dark matter halo (\citealt{nfw1997}, with $M_{\text{NFW}}=1.5\times 10^{13}$ M$_{\odot}$, $R_{\text{NFW}}=36.46$ kpc and $c_{\text{NFW}}=5$), a central dominant elliptical galaxy modelled with a Hernquist profile (\citealt{hernquist1990}, $M_{\text{cD}}=1.4\times10^{11}$ M$_{\odot}$ and $R_{\text{cD}}=10$ kpc), and a central SMBH with mass $M_{\bullet}=2.8\times10^8$ M$_{\odot}$ \citep[e.g.][]{CammelliMonaco2025,PianaPu2025}. The gas entropy profile follows a power-law $K(r) = K_{0} + K_{100} \left ( r/ 100 \text{ kpc} \right )^{\alpha_K}$ motivated by the ACCEPT cluster database \citep{cavagnolo2009}, rescaled to represent a galaxy group environment using $K_{0}=0.34$ KeV\,cm$^2$, $K_{100}=71$ KeV cm$^2$ and $\alpha_K=0.5$. The density and temperature profiles are obtained assuming hydrostatic equilibrium and are consistent with observed IGrM properties (see B26a). The central SMBH is modelled using a spherical sink region of radius $r_{\rm sink}=4\,\Delta x_{\min}\simeq0.4$ pc. At each timestep, gas within the sink radius is removed by resetting its density, temperature, and velocity to $\rho_{\rm sink} = 10^{-30} \ \mathrm{g \ cm^{-3}}$, $T_{\rm sink} = 1 \ \mathrm{K}$, and $v_{\rm sink} = 0$ km s$^{-1}$, preventing unresolved gas accumulation at the resolution limit. $M_{\bullet}$ is fixed throughout the simulation, and the gravitational softening length $\epsilon$ is chosen to be $\epsilon < r_{\rm sink}$. The total mass accreted over the simulated time span is negligible compared to the SMBH mass and does not affect the gas dynamics.

Radiative cooling is implemented using a temperature-dependent cooling function $\Lambda(T)$. For $T\ge10^{4.2}$ K, We adopt the cooling curve of \citet{schure2009}, appropriate for optically thin plasma in collisional ionisation equilibrium, assuming solar metallicity, consistent with an enriched hot central group atmosphere \citep[e.g.][]{Sun2012}. At lower temperatures, additional atomic and molecular cooling processes are included following \citet{gaspari2017}

\begin{align}\label{lambda}
\Lambda_{\text{cold}} &= 2 \times 10^{-19} 
   \exp[-1.184\times 10^5/(T+10^3)] \notag \\
&\quad + 2.8 \times 10^{-27}\sqrt{T}\,\exp[-92/T],
\end{align}
which approximates the combined contribution of several chemical processes, including atomic line cooling (hydrogen Ly$\alpha$, C$^{+}$, \ion{O}{I}, Fe$^{+}$ and Si$^{+}$), rovibrational cooling from molecules such as H$_2$ and CO, and gas--grain collisional cooling \citep{Koyama2000}, allowing gas to cool down to $\sim10$--$100$ K. Cooling losses are integrated using an exact implicit scheme \citep{Townsend2009,gaspari2012}, which is unconditionally stable and preserves positive internal energy.

In this work, we do not explicitly model AGN jets or winds. Instead, we adopt an idealized turbulence-driving scheme intended to represent the effective stirring of the hot atmosphere. Companion {\sc BlackHoleWeather} studies \citep[][C26a,b and P26a,b]{Cammelli2026a,Cammelli2026b, Piana2026a,Piana2026b} extend this framework to accretion-powered kinetic jets, indicating that feedback does not erase the CCA rain cycle but reshapes it by opening hot channels and shifting condensation towards the jet--ambient interface. Thus, AGN feedback is expected to modify the geometry and duty cycle of CCA, rather than fully suppress multiphase accretion.

Turbulence is sustained by a stochastic acceleration field evolved as an Ornstein--Uhlenbeck process in Fourier space \citep[e.g.][]{schmidt2009,grete2018,grete2025}. The forcing is solenoidal and injected at an intermediate spatial scale corresponding to a characteristic dimensionless mode number $n_{\rm peak}=4$, i.e.
$L_{\rm inj}=L_{\rm box}/n_{\rm peak}\simeq12.5$ kpc. This scale is not meant to correspond uniquely to a single physical driver. Rather, it is broadly consistent with the effective stirring expected from AGN feedback and/or sloshing and minor mergers \citep[e.g.][]{gaspari2012,vazza2012,zuhone2013,iapichino2017}. The driving strength is controlled by the prescribed root-mean-square acceleration amplitude $a_{\rm rms}$ and a correlation time $t_{\rm corr}=30$ Myr. Two turbulence regimes are considered (weak and strong) differing only in $a_{\rm rms}$. This controlled approach injects kinetic energy on large scales and allows it to cascade non-linearly to smaller scales. In this way, the flow self-consistently develops a turbulent spectrum representative of the stirred hot atmospheres of galaxy groups.
Because the acceleration field is applied in a volume-filling manner, the resulting phase coupling should not be interpreted as a literal statement that hot, warm, and cold gas respond identically to a given physical driver. Rather, the forcing maintains a stirred hot atmosphere from which colder phases condense and inherit part of the ambient kinematic structure, while their subsequent evolution is shaped by gravity, radiative cooling, cloud--cloud interactions, and multiphase mixing.

All simulations share identical initial conditions and differ only in the inclusion of radiative cooling and the strength of turbulent driving (Table~\ref{models}). The simulation suite comprises an adiabatic reference run, a cooling-only run, turbulence-only runs with different stirring levels, and simulations including both cooling and turbulence. The main CCA runs have cooling activated after 35 Myr with weak (\lowc) and strong (\highc) turbulence, driven with the same RMS accelerations adopted in the turbulence-only runs, i.e.~$a_{\rm rms}=6.2\times10^{-9}\,\mathrm{cm\,s^{-2}}$ (weak) and $a_{\rm rms}\simeq1.6\times10^{-8}\,\mathrm{cm\,s^{-2}}$ (strong). 
Over the simulated time, we target subsonic turbulence bracketing two extremal 3D Mach numbers of $\mathcal{M}\sim0.15$ in \lowc\ and $\mathcal{M}\sim0.4$ in \highc\ (as found by direct or indirect observations; e.g. \citealt{gasparichurazov2013,hofmann2016,xrism2025}; B26a), with corresponding velocity dispersions of $\sigma_v \simeq 60$–$90~\mathrm{km~s^{-1}}$ and $\sigma_v \simeq 210$–$230~\mathrm{km~s^{-1}}$, respectively. As a useful estimate, the turbulent energy injection rate per unit mass can be written as $\dot e_{\rm turb}\sim\sigma_v^3/L_{\rm inj}$. Using the measured velocity dispersions and the common injection scale $L_{\rm inj}\simeq12.5$
kpc, we obtain $\dot e_{\rm turb}\sim10^{-2}$ and
$\sim3\times10^{-1}\ {\rm erg\,g^{-1}\,s^{-1}}$ for \lowc\ and \highc, respectively, corresponding to a factor of $\sim25$ difference between the two runs. However, the associated turbulent-heating time scales as $t_{\rm heat}\sim e_{\rm th}/\dot e_{\rm turb}\sim t_{\rm eddy}\mathcal{M}^{-2}$ \citep[e.g.][]{gaspari2014}, implying that, for
the subsonic turbulence considered here, turbulent dissipation acts over many
eddy times. Since condensation usually occurs when
$t_{\rm cool}\sim t_{\rm eddy}$ (see Section~\ref{cca_diag}), cooling proceeds
faster than turbulent heating can offset it\footnote{For instance, using the characteristic Mach numbers of our two fiducial runs,
the scaling $t_{\rm heat}\sim t_{\rm eddy}\mathcal{M}^{-2}$ implies
$t_{\rm heat}\sim45\,t_{\rm eddy}$ for \lowc\ ($\mathcal{M}\sim0.15$) and
$t_{\rm heat}\sim6\,t_{\rm eddy}$ for \highc\ ($\mathcal{M}\sim0.4$).}

\begin{table}
    \centering
    \caption{Summary of the simulation suite.}
    \label{models}
    \begin{tabular}{lccc}
        \toprule
        \textbf{Simulation Name} & cooling & turbulence \\
        \midrule
        \bondi  & no & no \\
        \cool & yes & no \\
        \turbolow  & no & weak \\
        \lowc  & yes & weak \\
        \turbohigh    & no & strong \\
        \highc    & yes & strong \\
        \bottomrule
    \end{tabular}
    \tablefoot{Summary of the simulation suite. Each run is labelled by its short name and characterised by the physical processes included: radiative cooling and turbulence driving (weak and strong turbulence correspond to $\mathcal{M}\sim0.15$ and 0.4, respectively).}
\end{table}

Building on the results presented in B26a, we now focus on the accretion and variability properties of CCA in the same set of simulations. In B26a, we showed that radiative cooling in a turbulent stratified atmosphere leads to the formation of a strongly multiphase medium, with cold filaments and clouds condensing out of the hot halo and raining towards the central SMBH. In Figure \ref{tempslice} we show a density projection at increasing times for \highc\ (top) and \lowc \ (bottom). The morphology and spatial extent of the multiphase gas depend sensitively on the turbulence strength, which produces very different `BH weathers'. The \highc\ case develops an extended, highly fragmented network of filament-dominated structures on kiloparsec scale (`stormy' weather) and contains a cold gas mass that is larger by $2-3$ orders of magnitude, depending on radius, compared to \lowc. In contrast, the \lowc\ case produces a more centrally concentrated and less fragmented cold component (`rainy' weather). These morphological differences highlight the key role of turbulence in redistributing multiphase gas during the condensation cascade. Strong stirring promotes the formation of long, radially extended filaments and sustains multiphase structure over meso to macro-scales, while weaker turbulence favours rapid condensation followed by the collapse of cold gas into a compact nuclear core. Importantly, both morphologies arise within the same underlying halo and cooling setup, emphasising that filament extent and geometry are not fixed properties of the atmosphere but depend sensitively on how turbulence shapes the spatial development of CCA.

\begin{table}[h!]
    \centering
    \caption{Thermal phase classification used in this work.}
    \label{tab:thermal_phases}
    \begin{tabular}{lc}
        \toprule
        \textbf{Phase} & \textbf{Temperature range} \\
        \midrule
        Cold molecular & $T < 2\times10^{2}\,\mathrm{K}$ \\
        Cold          & $2\times10^{2} \le T < 1.6\times10^{4}\,\mathrm{K}$ \\
        Warm          & $1.6\times10^{4} \le T < 1.16\times10^{6}\,\mathrm{K}$ \\
        Hot (soft X-ray) & $1.16\times10^{6} \le T < 5.8\times10^{6}\,\mathrm{K}$ \\
        Hot (hard X-ray) & $T \ge 5.8\times10^{6}\,\mathrm{K}$ \\
        \bottomrule
    \end{tabular}
\tablefoot{The temperature intervals define the thermal phase bins used throughout the paper. The labels in parentheses indicate the approximate observational band in which each phase predominantly emits.}

\end{table}

Here, we investigate how these distinct multiphase weathers translate into differences in inflow variability and kinematical properties. Because stronger turbulent stirring enhances mixing and provides additional non-thermal support, the hot atmosphere in \highc\ requires a longer time to cool and condense than in \lowc, causing the two simulations to evolve on intrinsically different physical timescales. To enable a direct comparison between the two regimes, we normalise time by the raining time $t_{\rm rain}$ (tied to the effective cooling time), defined as the moment when the first cold ($T \approx 10^{4}$~K) gas begins to rain towards the centre. The raining time is measured starting from the moment when radiative cooling is switched on. Each simulation evolves up to $t/t_{\rm rain}=3$, where $t_{\rm rain}\simeq30$~Myr for \highc\ and $t_{\rm rain}\simeq7$~Myr for \lowc\ (correlated with the very inner cooling times).

For the analysis, we divide the gas into five thermal phases based on temperature: cold molecular, cold, warm, and hot gas further separated into soft and hard X-ray components. The temperature ranges defining each phase are listed in Table~\ref{tab:thermal_phases}.

To characterise the multiscale development of condensation and inflow, we partition the computational domain into four radial regimes: micro ($r \leq 0.1~\mathrm{kpc}$), meso ($0.1 < r \leq 1~\mathrm{kpc}$), inner macro ($1 < r \leq 10~\mathrm{kpc}$), and outer macro ($r > 10~\mathrm{kpc}$). This scale separation enables a consistent comparison of gas properties and accretion dynamics from kpc scales down to the central parsec, which is possible owing to the very high resolution reached in our simulations ($\approx 0.1$ pc). This multi-phase and multi-scale decomposition provides the common framework used throughout the paper to link the thermodynamical state of the gas to the variability, kinematics, and efficiency of SMBH fueling.

\section{Multiphase mass inflows}\label{sec:accretion}

To assess how the thermodynamical state of the IGrM affects SMBH fueling, we analyse the time evolution of the mass inflow rates in our simulations.

\subsection{SMBH accretion}\label{sec:smbhaccretion}
\begin{figure}
\centering
\includegraphics[width=0.95\columnwidth]{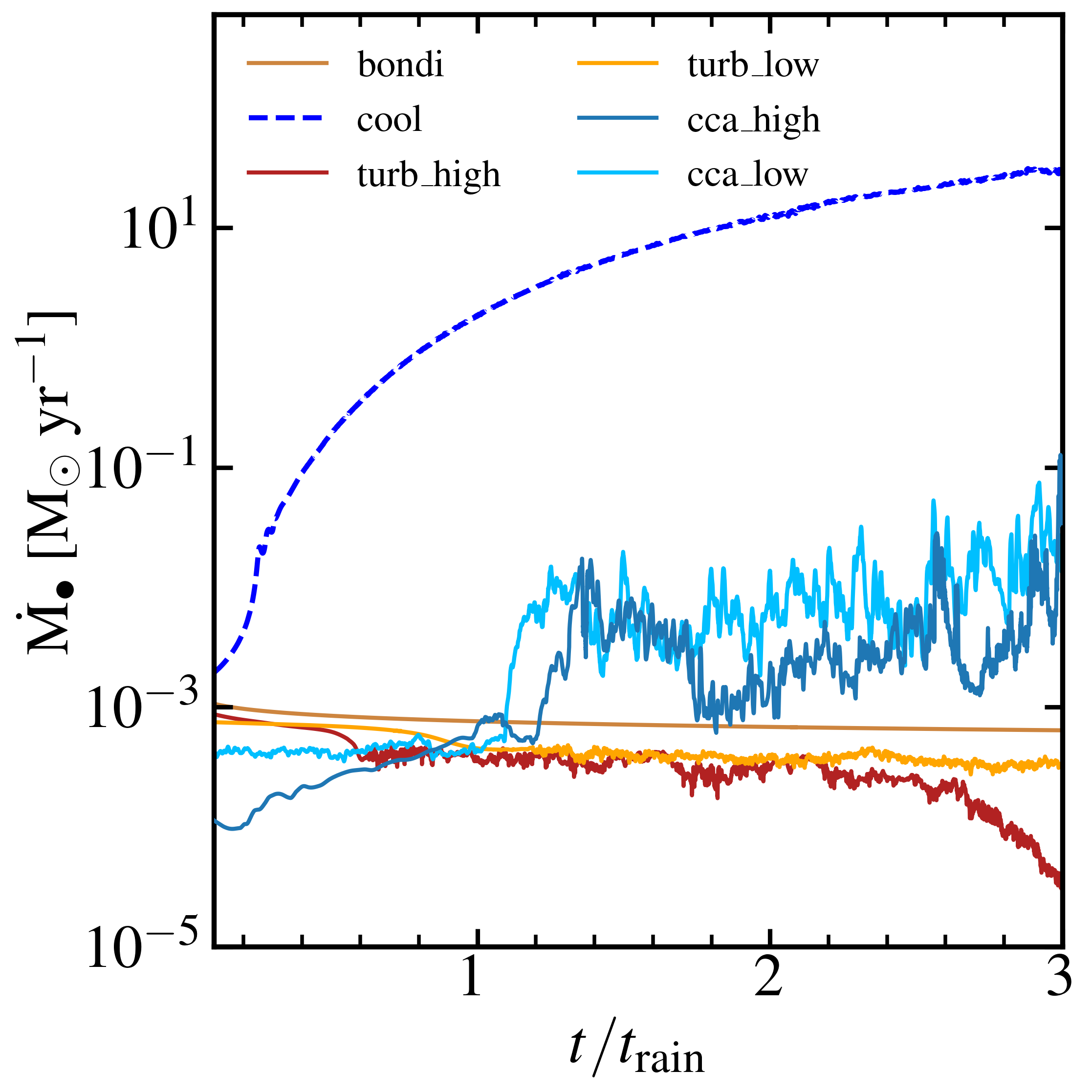} 

\caption{SMBH accretion rate $\dot{M_\bullet}$ as a function of normalised time $t/t_{\text{rain}}$ for the \highc\ (blue line, stormy weather) and the \lowc\ (light blue line, rainy weather) simulations, compared with the only turbulence simulations (sunny weather) \turbohigh\ (red line) and \turbolow\ (orange line), with the radiative cooling simulation \cool\ (blue dashed line) and with the idealized adiabatic simulation \bondi \ (brown line). The CCA phenomenon produces a high variability accretion rate due to its chaotic nature.}\label{accretionlowc}

\end{figure}

We define the SMBH accretion rate as the mass accreted within the sink region ($r<0.4$ pc) during each simulation timestep. Figure \ref{accretionlowc} shows the resulting accretion histories, normalised by $t_{\text{rain}}$, for \highc\ (blue) and \lowc\ (light blue), compared with the adiabatic (\bondi, brown), and purely turbulent control runs \turbohigh\ (red) and \turbolow\ (orange). The figure highlights both the high variability and the contrasting fueling behaviour across different turbulent regimes.

The classic \citet{bondi1952} accretion rate is
\begin{equation}
\dot{M}_{\rm B}
= 4\pi\,\lambda(\gamma)\,\frac{(G M_\bullet)^2\,\rho_\infty}{c_{s,\infty}^{\,3}} \, ,
\end{equation}

where $\lambda$ is a dimensionless factor (of order unity) that depends on the adiabatic index $\gamma$,
\begin{equation}
\lambda(\gamma)
= \left(\frac{1}{2}\right)^{\frac{\gamma+1}{2(\gamma-1)}}
\left(\frac{5-3\gamma}{4}\right)^{-\frac{5-3\gamma}{2(\gamma-1)}} \, ,
\end{equation}

where $\rho_\infty$ and $c_{s,\infty}$ denote the ambient density and sound speed at large radii.
This convenient expression has been widely employed in the past years across a broad range of astrophysical contexts. In particular, a lot of theoretical, numerical, and observational studies assume that the accretion flow follows the Bondi solution even when the Bondi radius 
($r_{\rm B}=G M_\bullet / c_{\rm s,\infty}^{2}$) is unresolved \citep[e.g.][]{DiMatteo2005,cattaneo2007,Booth2009,yang2012}. However, realistic galactic atmospheres deviate from the highly idealized and restrictive assumptions of a homogeneous, adiabatic, radial and steady flow, requiring numerical models that incorporate stratification, cooling, and turbulence.
Thus, while the adiabatic \bondi\ simulation provides a controlled reference for the hot-mode accretion regime, it cannot be used to infer a realistic SMBH accretion rate. After a brief initial transient, the accretion rate settles to a nearly constant value of $\dot{M}_\bullet\simeq6\times10^{-4}\,\mathrm{M}_\odot\,\mathrm{yr^{-1}}$, in close agreement with the analytic Bondi prediction based on the ambient hot-gas density and temperature. The absence of cooling and turbulence preserves the self-similar inflow structure, resulting in a smooth, steady accretion pattern without significant variability. Minor deviations from the analytic rate (at the $\sim$30\% level) arise over time from the stratification of the galactic atmosphere in the inner 50 kpc, which slightly lowers the central density relative to the idealized uniform case. Overall, the simulation confirms that, under realistic thermodynamic gradients, purely adiabatic Bondi accretion remains laminar, quasi-spherical, and dynamically stable over multiple dynamical times. As an additional check, we also computed the instantaneous Bondi rate using the time-dependent hot-gas properties measured in the fiducial simulations. These estimates are typically \(\sim0.3\)--\(0.4\) dex below the accretion rate of the adiabatic \bondi\ run.

The effect of subsonic turbulence in the absence of radiative cooling is isolated in the purely turbulent runs. The \turbolow\ simulation features moderate solenoidal stirring (\(\mathcal{M} \simeq 0.1\text{--}0.2\)), where the turbulent kinetic energy constitutes only a minor fraction of the thermal budget, yielding a mildly overheated halo as might follow an AGN outburst or minor merger. The flow is in a pure hot accretion mode, leading to a reduced accretion rate compared to the adiabatic case (\bondi): the accretion rate $\dot{M}_\bullet$ decreases by a factor of a few below the Bondi reference, settling around $\dot{M}_{\bullet} \simeq (3$--$4) \times 10^{-4}\,\mathrm{M}_\odot\,\mathrm{yr^{-1}} \simeq 0.6\,\dot{M}_{\mathrm{B}}$. The inflow becomes increasingly intermittent, with fluctuations driven by the formation of vortical structures that redistribute angular momentum. The reduction in accretion occurs because turbulent motions generate vorticity in the stratified halo, producing small rotating eddies that hinder the direct inflow of gas towards the black hole. These eddies provide effective angular-momentum support, forcing the gas to circulate rather than accrete. In addition, transient bulk motions induced by turbulence create a non-zero relative velocity between the black hole and the surrounding gas, further decreasing the capture rate in a Bondi–Hoyle–like fashion \citep{bondi1944}. 

The \turbohigh\ run, characterised by stronger subsonic stirring ($\mathcal{M} \simeq 0.4$), exhibits a similar evolution but with a slightly lower accretion rate of $\dot{M}_{\bullet} \simeq (2$--$3) \times 10^{-4}\,\mathrm{M}_\odot\,\mathrm{yr^{-1}} \simeq 0.5\,\dot{M}_{\mathrm{B}}$. In this case, turbulent eddies are produced more rapidly and the maximum velocity dispersion is reached sooner, as the eddy turnover time is shorter (Section \ref{cca_diag}). 
Towards the end of the simulation ($t \approx 100~\mathrm{Myr}$), the accretion rate declines further, dropping below $10^{-4}\,\mathrm{M}_\odot\,\mathrm{yr^{-1}}$. This drop coincides with a rise in the velocity dispersion (due to the more diffuse stirred atmosphere), which increases turbulent support and effectively makes the gas too kinematically hot to be accreted efficiently onto the SMBH.

The \cool\ simulation illustrates the opposite limiting case, in which radiative cooling is included but the gas is not perturbed by turbulence. In this run, the central gas cools efficiently, loses pressure support, and accumulates towards the sink region (akin to a pure `cooling flow'), leading to a rapid and sustained increase of the SMBH accretion rate up to $\dot{M}_\bullet \simeq 10-20\,\mathrm{M}_\odot\,\mathrm{yr^{-1}}$

As found in B26a, including radiative cooling fundamentally changes the accretion mode: multiphase condensation sets in and the SMBH transitions from `sunny' weather (i.e. a hot turbulence dominated atmosphere) to CCA precipitation (e.g. rainy or stormy weathers).
The formation of cold filaments and clumps characteristic via CCA, followed by their infall towards the centre, boosts the accretion rate by orders of magnitude. Despite their contrasting filamentary morphologies (see Figure \ref{tempslice}), the two main runs \highc\ and \lowc\ display accretion histories similar in shape.

In the \highc\ simulation, the accretion rate is highly variable, spanning 1--2 orders of magnitude ($10^{-3}$--$10^{-1}\,\mathrm{M}_\odot\,\mathrm{yr^{-1}}$). At early times, the accretion remains comparable to the purely turbulent counterpart \turbohigh. Following the first major condensation event around $t/t_{\mathrm{rain}} \sim 1$, the inflow becomes bursty and irregular, with $\dot{M}_\bullet$ oscillating between minima of $\approx 10^{-3}\,\mathrm{M}_\odot\,\mathrm{yr^{-1}}$ and peaks up to $\sim 10^{-1}\,\mathrm{M}_\odot\,\mathrm{yr^{-1}}$. These rapid fluctuations are driven by interactions among the infalling cold clouds, which can then lose angular momentum and fall into the centre. 
In this regime, the large-scale condensation remains extended and filamentary, but the innermost feeding does not settle into a single stable, coherent filamentary channel; instead, accretion proceeds through recurrent, clumpy rain events driven by cloud and filament interactions.

The \lowc\ simulation yields a slightly higher average inflow than \highc, as weaker turbulent velocity dispersion allows denser and more coherent filaments to reach the SMBH more efficiently. At early times ($t < t_{\text{rain}}$), the accretion rate remains comparable to the \turbolow\ case, with $\dot{M}_\bullet \lesssim 10^{-3}\,\mathrm{M}_\odot\,\mathrm{yr^{-1}}$ during the first $t_{\text{rain}}=7$ Myr. Once condensation sets in, $\dot{M}_\bullet$ rises sharply by more than an order of magnitude, reaching $\sim 10^{-2}\,\mathrm{M}_\odot\,\mathrm{yr^{-1}}$, after which it fluctuates around this level with high variability, with a shape similar to \highc. The overall behaviour reflects a chaotic, filamentary mode of condensation. On average, both \highc\ and \lowc\ accretion rates remain strongly super-Bondi.

\begin{figure*}
\centering
\includegraphics[width=0.88\textwidth]{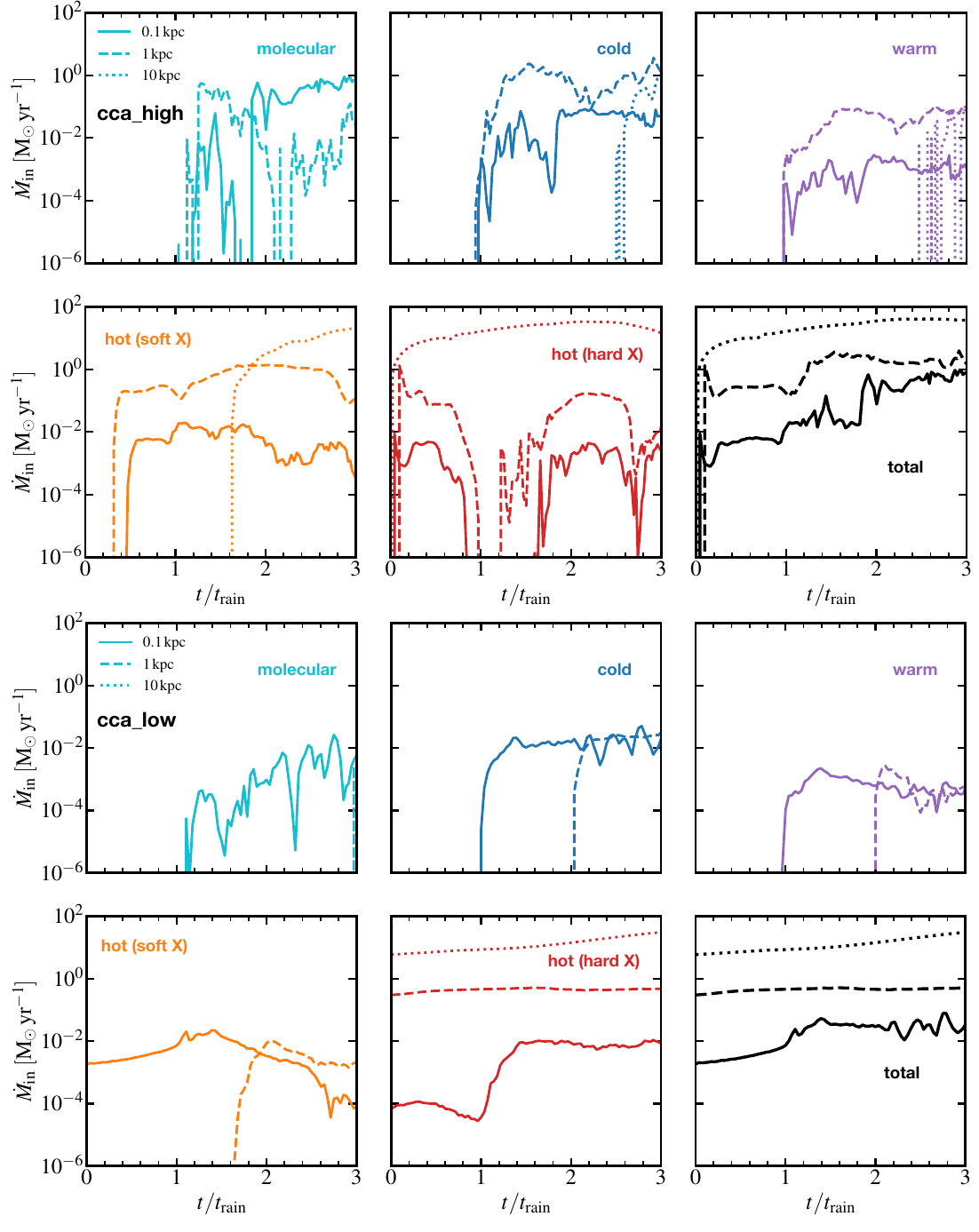} 

\caption{Mass inflow rates at different radii. The mass inflow rate $\dot{M_{\rm in}}$ is shown as a function of normalised time $t/t_{\rm rain}$ for different thermal phases (panels) and measured at spherical radii of $r = 0.1$~kpc (solid lines), $1$~kpc (dashed lines), and $10$~kpc (dotted lines).
The top two rows correspond to the \highc\ simulation, while the bottom two rows show the \lowc\ case. The last panel displays the total inflow rate for both simulations (black), obtained by summing over all thermal phases.}\label{inflowradius}

\end{figure*}

\subsection{Mass inflow rates}

Having characterised the accretion rate onto the SMBH, we now investigate how gas is transported through the halo at progressively larger radii. Figure~\ref{inflowradius} shows the mass inflow rates as a function of $t/t_{\rm rain}$ for \highc\ (top panels) and \lowc\ (bottom panels) measured at $r=0.1$~kpc (solid lines), $1$~kpc (dashed lines), and $10$~kpc (dotted lines). We compute the mass inflow rate through a spherical shell at radius $r$ by summing the radial mass flux over inflowing cells only:
\begin{equation}
\dot{M_{\rm in}}(r)
=
-\sum_{v_{r,i}<0} \rho_i\, v_{r,i}\,
\frac{V_i}{2\,\Delta r} \, ,
\end{equation}
where $\rho_i$ is the gas density, $v_{r,i}$ the radial velocity, $V_i$ the cell volume, and $\Delta r=0.03$ kpc is the half--thickness of the shell. The sum includes all cells with $|r_i-r|<\Delta r$ and $v_{r,i}<0$, so that $\dot{M_{\rm{in}}}(r)>0$ by construction denotes inward mass flux. Because turbulent and bulk motions can produce large, partially cancelling inward and outward streams, $\dot{M_{\rm{in}}}(r)$ should be interpreted as an instantaneous shell inflow rate, rather than as a global mass deposition (cooling-flow) rate.

The total inflow rate (black line) exhibits a clear radial hierarchy in both simulations. At large radii ($r=10$ kpc), the inflow is dominated by the hot phase and reflects the global inward movement of the atmosphere. The mass inflow rate at this scale reaches values between $10$–$50\ \mathrm{M_\odot\ yr^{-1}}$, with $\dot{M_{\rm{in}}}$ in \highc\ increasing more rapidly than in \lowc. Moving towards smaller radii, the inflow progressively decreases, indicating that only a fraction of the large-scale hot inflow effectively propagates to the central region. At $r\approx1$ and $0.1$ kpc, the inflow becomes increasingly intermittent in both simulations, reflecting the transition to a multiphase regime. 
This radial meso-scale transition marks the onset of local multiphase condensation: the hot inflow sets the outer boundary condition, while the inner kpc becomes regulated by phase conversion, fragmentation, and cloud interactions.
In \highc, the variability strongly amplifies towards small scales, with pronounced bursty events, particularly for $t/t_{\rm rain} > 2$, coincident with the formation of an extended network of cold filaments. These structures fragment and collide, producing rapid fluctuations in the inflow rate. 
In contrast, \lowc\ maintains a lower and smoother inflow across most radii. The hot phase dominates the large-scale inflow, while colder components become significant mainly inside $\sim1$ kpc, with the molecular phase largely confined to $r=0.1$ kpc.
The reduced turbulence level leads to a more compact cold gas distribution and consequently to a less fragmented and less variable inflow towards the centre.

Following this global radial hierarchy, we now examine the phase decomposition of the inflow. After $t=t_{\rm rain}$, the mass transport becomes strongly multiphase in both simulations.
In \highc, molecular gas rapidly becomes the dominant contributor at $r=0.1~\mathrm{kpc}$, displaying strong variability. Around $t\sim2 \ t_{\rm rain}$, a substantial molecular reservoir ($\sim10^{6} \ {\rm M}_\odot$ at meso-scales) assembles and subsequently rains towards the nucleus, producing molecular inflow rates approaching $\sim1~\mathrm{M_\odot \ yr^{-1}}$. The fact that molecular inflow is already significant at $r=1~\mathrm{kpc}$ (between $\sim10^{-3}~\mathrm{M_\odot \ yr^{-1}}$ and $\sim0.5~\mathrm{M_\odot \ yr^{-1}}$, with a strong variability) but negligible at $10~\mathrm{kpc}$ supports a scenario in which condensation occurs in situ at intermediate radii, followed by chaotic inflows. Cold inflow rates are higher at $r=1$ kpc, reaching $\gtrsim 1~\mathrm{M_\odot \ yr^{-1}}$, whereas they follow a trend similar to molecular gas at $r=0.1$ kpc, increasing after $t=2t_{\rm rain}$. 
At $r=10$ kpc cold inflow rates are non negligible after $t=2.5t_{\rm rain}$.
The cold and warm phases evolve coherently, reflecting their thermodynamic coupling at phase interfaces, as also discussed in B26a: warm gas is typically formed at the boundaries between cold clumps/filaments and the surrounding hot plasma.

\begin{figure}
\centering
\includegraphics[width=0.95\columnwidth]{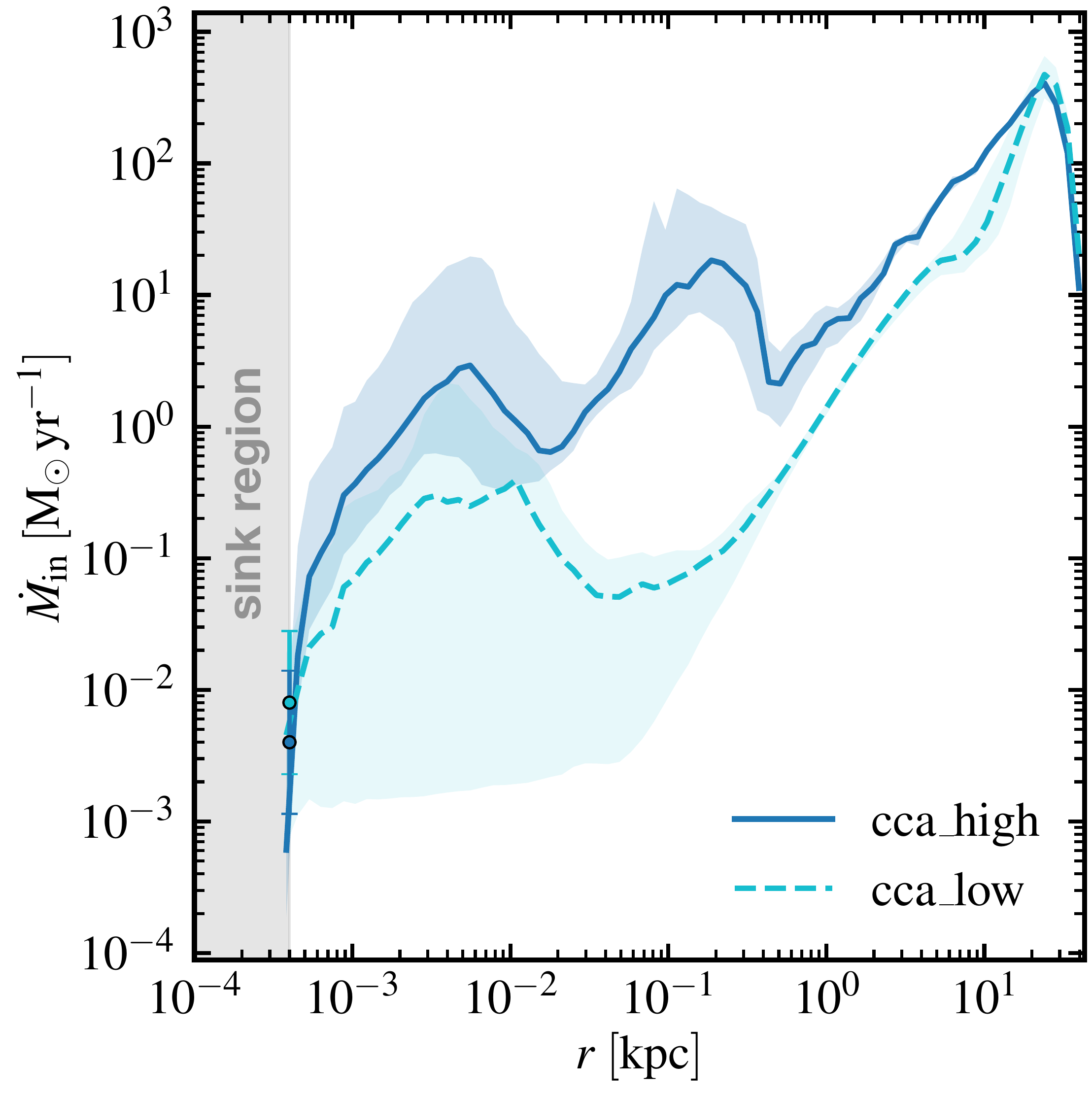} 

\caption{Mass inflow rate as a function of radius for \highc\ (solid line) and \lowc\ (dashed line). The lines represent the average value over the full simulation, while the shaded regions indicate the $1\sigma$ temporal dispersion. The grey shaded area marks the sink region ($r<4\times10^{-4}$ kpc). The points represent the average SMBH accretion rate $\dot{M}_{\bullet}$ measured on the sink, with error bars showing the $1\sigma$ variability.}\label{mdot_radial}

\end{figure}

By contrast, in \lowc\ the molecular component remains much weaker ($\sim10^{-2}~\mathrm{M_\odot \ yr^{-1}}$) and largely confined to the innermost region. The cold and warm have similar values at $0.1$  and $1$ kpc with the warm phase having smaller inflow rates, but the overall contribution remains modest compared to \highc. The reduced turbulence suppresses large-scale fragmentation and limits the buildup of an extended molecular reservoir. 

Figure \ref{inflowradius} also reveals a clear thermodynamic and radial ordering of the inflow. In both simulations, the hot phase dominates the transport at large radii, setting the large-scale supply background, whereas progressively colder phases emerge only towards smaller radii and later times. In \highc, the fact that molecular inflow is already substantial at $r=1$ kpc but remains negligible at $r=10$ kpc indicates that the coldest gas does not simply advect inward from the outer halo, but is assembled in situ at intermediate radii through a condensation cascade. The coherent evolution of the warm and cold components further supports this interpretation, suggesting that they trace phase interfaces and transitional cooling layers, while the molecular phase becomes dominant only after a sufficiently dense inner reservoir has formed. Compared with this stormy, strongly intermittent behaviour, \lowc\ displays a more radially confined and smoother transport mode, in which multiphase inflow remains modest outside the central kpc and the molecular component is largely restricted to the innermost region.

These inflow rates underscore a pronounced divergence in how cold gas is transported across kpc and meso-scales in the two turbulent weathers. Yet, despite order-of-magnitude differences in multiphase inflow at larger radii, the SMBH accretion rates are remarkably similar. This implies that, in these simulations, SMBH feeding is not determined solely by the instantaneous cold-gas supply at kiloparsec and/or macro-scales. Instead, the accretion rate is also shaped locally within the meso-scale, where angular-momentum cancellation and cloud–cloud interactions regulate how efficiently the inflowing material can be captured. In this sense, distinct large-scale halo-weather states, stormy in \highc\ and rainy in \lowc, can converge towards comparable micro-scale accretion efficiencies.

Figure \ref{mdot_radial} shows the average total mass inflow rate as a function of radius for both fiducial simulations. 
At large radii ($r\gtrsim10$ kpc), the two profiles are broadly comparable within the $1\sigma$ scatter, since both simulations start from the same hot-halo initial conditions and condensation has only a limited impact at these distances.
At $r<10$ kpc \highc\ shows a higher inflow rate at all radii. A prominent peak is present between $r\approx0.1$ kpc and $r\approx1$ kpc, corresponding to the meso-scale region where multiphase gas is particularly present. This feature is present only in \highc\ and could be used as a diagnostic to differentiate between the stormy and rainy regimes.
In \highc, the cold gas extends from several kpc down to the central region. The resulting extended and fragmented multiphase atmosphere enhances the inflows, generating a pronounced bump of about one order of magnitude ($\approx 5$--$6$ M$_{\odot}$ yr$^{-1}$). This peak in $\dot{M}_{\rm in}$ arises because stronger turbulent stirring in \highc\ builds an extended meso-scale reservoir of cold/warm clouds and filaments, which is then efficiently channelled inward. In contrast, \lowc\ develops a more centrally concentrated cold component, lacking the extended reservoir seen in \highc\ (Figure~\ref{tempslice}; see also B26a for the morphological analysis). As a result, the inflow enhancement between $r\approx1$ and $0.1$ kpc is absent in \lowc.

At the innermost resolved radii ($r\sim10^{-3}$--$10^{-2}$ kpc), both simulations display a secondary bump associated with micro-scale accretion. This similarity between the two simulations shows that \highc\ is not purely \textit{stormy} at all scales. While strong turbulence creates a fragmented and extended multiphase atmosphere at kpc and meso-scales, the character of the flow changes as gas moves inward. At micro-scales, the inflow naturally reorganises into a more centrally concentrated configuration, effectively transitioning from stormy large-scale weather to rainy inner accretion. At $r<10^{-3}$ kpc the average inflow rates in the two simulations are of the same order of magnitude ($6\times 10^{-3}$ M$_{\odot}$ yr$^{-1}$ for \lowc\ and $2\times 10^{-2}$ M$_{\odot}$ yr$^{-1}$ for \highc). In \lowc\ the inflow rate shows large variability and a strongly skewed distribution. In contrast, in \highc\ the distribution is less skewed, suggesting a more continuous and steadier feeding mode. While the average accretion level is similar, \highc\ provides a more consistent supply of gas to the SMBH, whereas \lowc\ is characterised by episodic and highly variable inflow.

Stormy and rainy weathers are not separate or mutually exclusive states, but scale-dependent expressions, in both space and time, of the same CCA process. Stormy weather mainly characterises the large-scale fragmentation and chaotic transport of multiphase gas, while rainy weather captures the final, more centrally concentrated phase of accretion. A similar convergence may also arise in simulations with major AGN jet episodes (C26a,b and P26a,b): even in the strongly stirred \highc\ run, the multiphase cascade ultimately approaches an inner feeding flow similar to \lowc. At the same time, this interpretation should be viewed as physically motivated rather than definitive, since the present simulations do not yet include self-regulated, time-dependent AGN feedback.

\subsection{Eddington ratios}

\begin{figure}
\centering
\includegraphics[width=0.95\columnwidth]{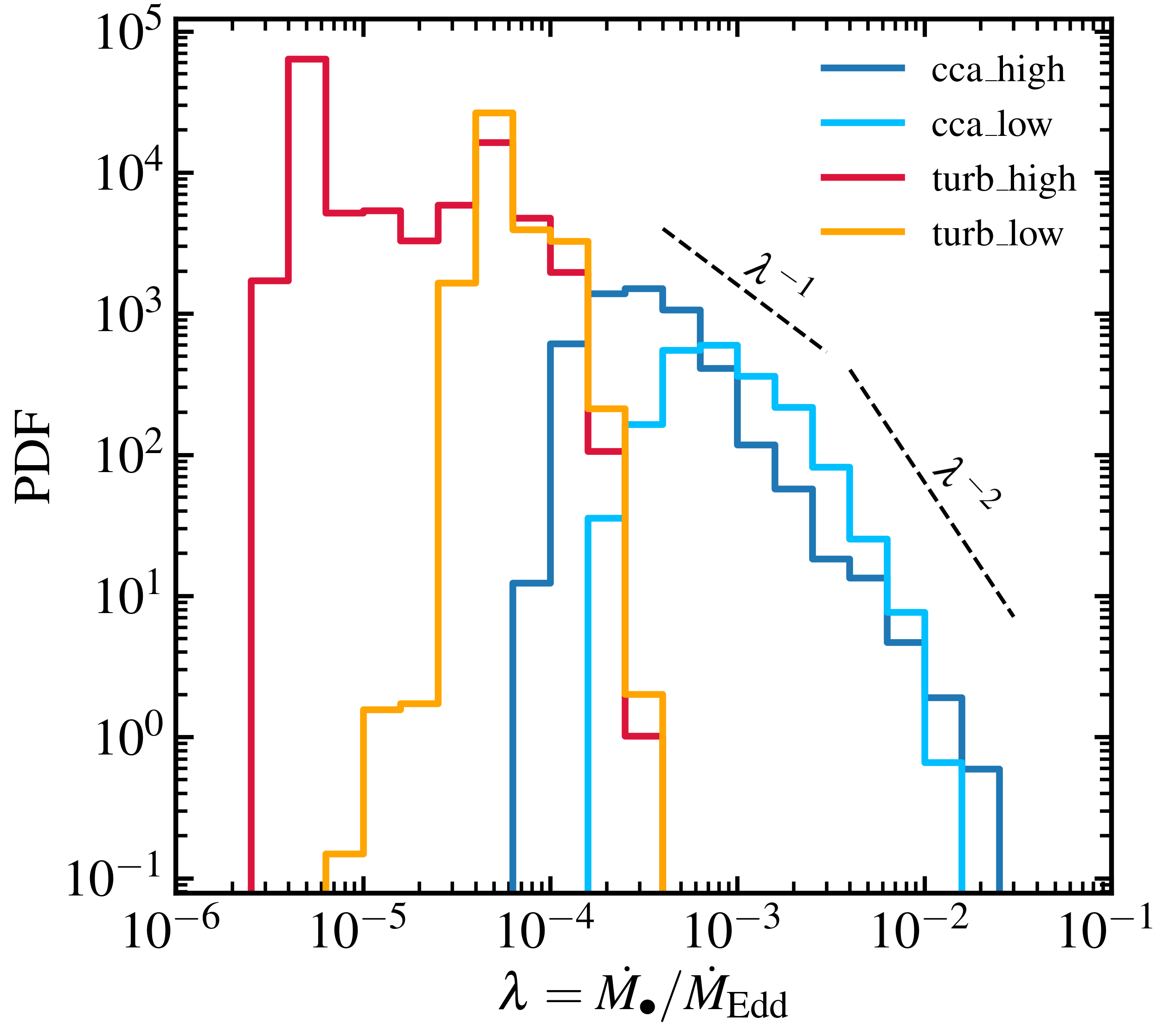} 

\caption{Probability density functions (PDFs) of the Eddington-normalised accretion rate, $\lambda = \dot{M_{\bullet}}/\dot{M}_{\rm Edd}$, for the \highc\ (blue), \lowc\ (light blue), \turbohigh\ (red) and \turbolow\ (orange).
The distributions of the fiducial CCA simulations peak at low Eddington ratios (a few $\times10^{-4}$), indicating that the SMBH spends most of its time in a low-accretion regime, while a high-$\lambda$ tail extends towards a self-similar power law.}\label{pdf_lambda}
\end{figure}

While the absolute accretion rates provide insight into the efficiency of
gas inflow onto the SMBH, their physical relevance is best assessed by comparing them to the Eddington accretion rate $\dot{M}_{\rm Edd}$, defined as the accretion rate required to produce the Eddington luminosity $L_{\rm Edd} = 4\pi G M_{\rm BH} m_p c / \sigma_T$, where $\sigma_T = 6.65\times10^{-25}\,{\rm cm}^2$ is the Thomson scattering cross section, $m_p = 1.67\times10^{-24}\,{\rm g}$ is the proton mass and $c = 3\times10^{10}\,{\rm cm\,s^{-1}}$ is the speed of light. Assuming a radiative efficiency $\eta=0.1$, we obtain $\dot{M}_{\rm Edd} = L_{\rm Edd}/(\eta c^2)=6.2$ M$_{\odot}$ yr$^{-1}$. Figure~\ref{pdf_lambda} shows the probability density functions (PDFs) of the Eddington-normalised accretion rate, $\lambda \equiv \dot{M_{\bullet}}/\dot{M}_{\rm Edd}$. By construction, $\lambda$ provides a dimensionless measure of the accretion rate relative to the black hole mass. Focusing on the turbulence-driven, hot-mode accretion runs, \turbohigh\ has an average Eddington ratio of $\lambda \approx 10^{-5}$, while \turbolow\ exhibits a slightly higher mean value, $\lambda \approx 5\times10^{-5}$.
Despite their similar average accretion levels, the two simulations display markedly different distributions: \turbohigh\ shows a broader spread, extending from $\lambda \sim 10^{-6}$ up to $\sim2\times10^{-4}$, whereas \turbolow\ is more narrowly clustered around its mean value. Both fiducial simulations, \highc\ and \lowc, exhibit accretion predominantly at low Eddington ratios, with distributions spanning $\lambda \sim 10^{-4}$--$10^{-2}$. Such values are expected for massive systems in low-redshift environments \citep[e.g.][]{RussellMcNamara2013}, where the SMBH has already grown to large masses and the surrounding IGrM is
predominantly hot, in the so-called radio mode feedback ($\lambda < 10^{-2}$). These low Eddington ratios do not imply negligible accretion in absolute terms; rather, they primarily reflect the large black hole mass, which sets a correspondingly high Eddington accretion rate. \highc\ peaks at $\lambda \approx 3\times10^{-4}$, whereas \lowc\ peaks at $\lambda \approx 7\times10^{-4}$, indicating a systematically higher accretion efficiency in the latter case, as highlighted also by Figure \ref{accretionlowc}. \highc\ is highly skewed towards large $\lambda$ , whereas \lowc\ exhibits a more Gaussian distribution. 
At higher Eddington ratios, both simulations PDFs exhibit an extended tail that roughly follows a power law with index between $-1$ and $-2$, as indicated by the dashed lines. In an idealized simulation with unlimited resolution, such a tail would be expected to extend to higher $\lambda$, albeit with progressively decreasing frequency, reaching $\lambda>10^{-2}$. Extending the integrations to Gyr timescales would also better sample the rare, high-accretion episodes that populate the tail, likely flattening the inferred slope and bringing it closer to unity.

Overall, the Eddington-ratio distributions indicate that the black hole spends most of its time in a low-accretion, maintenance-mode state, with growth driven by intermittent, chaotic infall of condensed cold clouds rather than by prolonged high-$\lambda$ episodes. High-$\lambda$ events are still expected, but they remain rare and transient.
This behaviour is broadly consistent with observations of radio-mode AGN in massive galaxies and cluster cores. Low-excitation radio galaxies (LERGs) typically accrete below $\sim 10^{-2}\,L_{\rm Edd}$, whereas high-excitation systems are more commonly found above this threshold \citep{BestHeckman2012}. Likewise, low-redshift BCGs hosting X-ray cavities are often strongly radiatively inefficient \citep{RussellMcNamara2013,HlavacekLarrondo2013}, with radiative Eddington ratios spanning the same broad low-$\lambda$ regime explored here, from quiescent sunny states to more active rainy/stormy phases. More generally, the local AGN population is dominated by low-Eddington accretion, while radiatively efficient, high-$\lambda$ systems constitute only a minor fraction of the total population \citep[e.g.][]{BestHeckman2012,AirdCoil2012,AirdCoil2018}. This is in good agreement with our low-redshift simulations, in which accretion at $\lambda>10^{-2}$ occurs for only $\sim1.5\%$ of the time in \highc\ and $\sim0.8\%$ in \lowc, confirming that the overall growth history is dominated by low-$\lambda$ phases. This interpretation is further supported by observational studies of cluster cores: \citet{Mcdonald2018} identify $\lambda\sim10^{-2}$ as the characteristic transition below which AGN remain in the low-accretion feedback regime, while \citet{Somboonpanyakul2022} show that Phoenix-like, high-accretion BCGs (quasar-like) are very rare and that most BCGs host AGN accreting at low relative-to-Eddington rates.

Importantly, this maintenance-mode regime does not imply purely hot, adiabatic feeding from the ambient halo. In the {\sc BlackHoleWeather} scenario, most of the accreted mass is instead supplied by cold gas that condenses out of the hot atmosphere through turbulence-driven thermal instability, namely CCA. In this sense, CCA provides a natural mechanism to reconcile predominantly low Eddington ratios with sustained, and at times super-Bondi, accretion in radio-mode systems. At the same time, nearby ellipticals and FR~I radio galaxies show that low-Eddington accretion can still sustain substantial jet power despite very low radiative efficiency \citep{Allen2006,Balmaverde2008}, reinforcing the interpretation that the simulated CCA inflow is primarily associated with kinetic/radio-mode feedback rather than persistent quasar-mode emission. More generally, CCA is expected to operate across a wide range of environments, including high-redshift, quasar-like systems: while the characteristic median accretion rate depends on ambient conditions, black-hole mass, and thermodynamic state, the underlying chaotic and scale-free nature of the feeding process should remain broadly self-similar \citep[see also][]{fiore2024}.

\begin{figure}
\centering
\includegraphics[width=0.95\columnwidth]{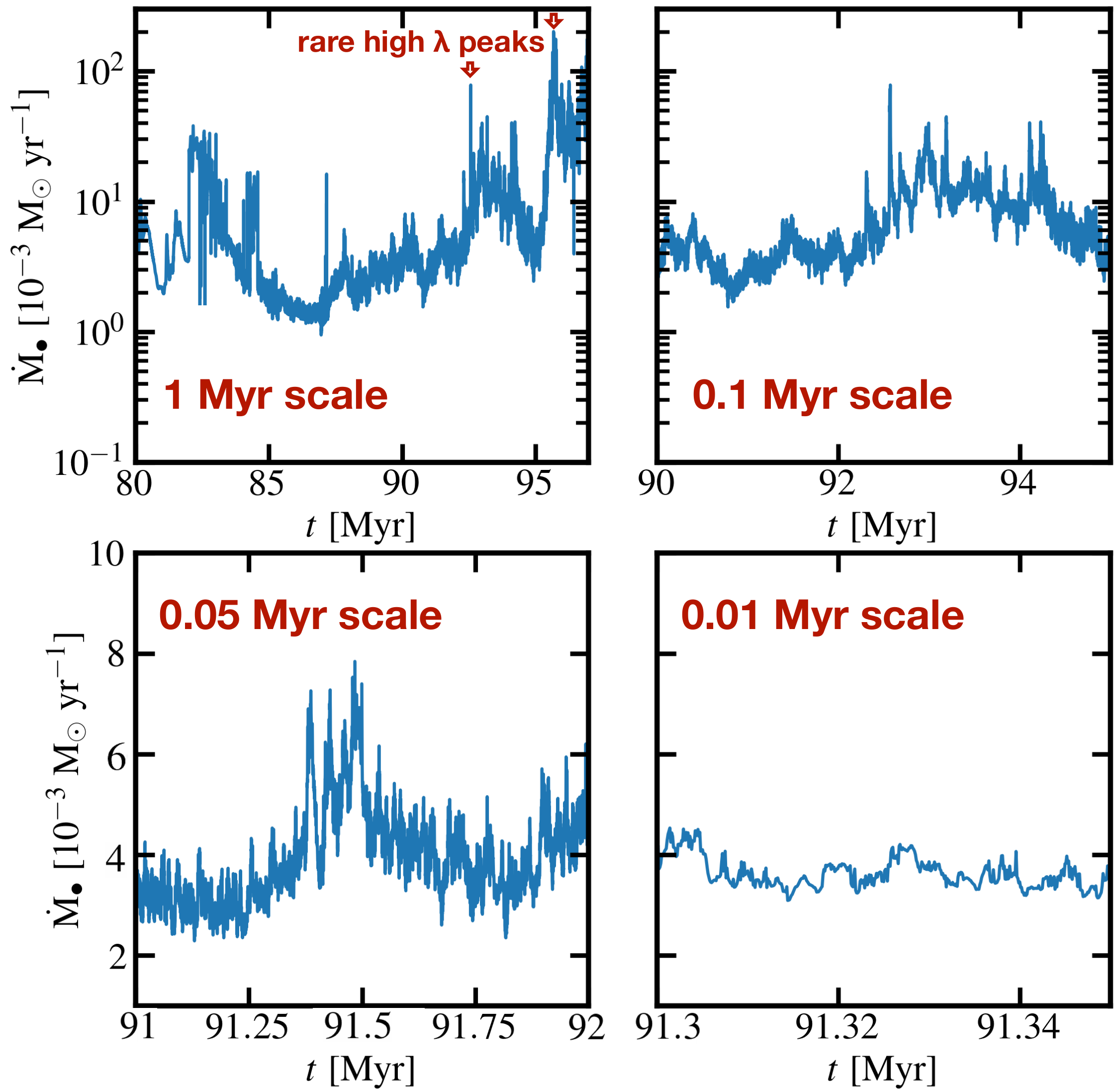} 

\caption{SMBH accretion rate $\dot{M_\bullet}$ as a function of time in the \highc\ simulation shown over progressively shorter time intervals. The top-left panel spans the time range $t= [80,97]$ Myr, while the top-right panel zooms into $t = [90,95]$ Myr. The bottom-left panel further magnifies the interval $t = [91,92]$ Myr, and the bottom-right panel shows a close-up of $t= [91.3,91.35]$ Myr. Variability persists across all resolved timescales, with its amplitude progressively decreasing towards shorter timescales. Two red arrows highlight rare high-$\lambda$ events.}\label{accretionzoom}

\end{figure}

\section{Time variability}\label{sec:timevariability}

A defining property of the SMBH accretion rate in both fiducial simulations is its strong time variability \citep{PaolilloPapadakis2025}. Both simulations show fluctuations spanning up to $\sim2$ orders of magnitude and flickering across a broad range of timescales. If transmitted through the unresolved inner accretion flow, such variability could induce substantial changes in the AGN power output, although, since the micro-scale accretion disc is not resolved, $\dot{M}_{\bullet}$ should be interpreted as the fueling rate into the unresolved inner region; once gas joins the disc, the actual SMBH accretion rate may depend on the disc properties and viscous timescale \citep{ShakuraSunyaev1973,TurnerReynolds2021}.

From Figure~\ref{accretionlowc}, it is evident that the accretion rates in both \highc\ and \lowc\ display a complex, hierarchical variability pattern. The accretion history exhibits a sequence of peaks nested within peaks across multiple timescales, consistent with the scale-dependent nature of CCA. This behaviour is illustrated in Figure~\ref{accretionzoom}, which shows a zoomed view of the accretion rate of the \highc\ run over different temporal intervals. At each timescale, the accretion rate exhibits several peaks likely associated with the infall and interactions of multiphase structures in the inner flow. These fluctuations occur over timescales ranging from $\sim10$~Myr down to $0.01$~Myr. The amplitude of variability depends strongly on the characteristic timescale: in the upper-left panel, fluctuations on $\sim1$~Myr scales reach amplitudes up to a factor of $\sim25$ (corresponding to $\sim1.4$~dex); in the upper-right panel, variations on $\sim0.1$~Myr scales reach a factor of $\sim3$ ($\sim0.5$~dex); in the lower-left panel, $\sim0.05$~Myr fluctuations have amplitudes of about $\sim75\%$ ($\sim0.25$~dex); and in the lower-right panel, $\sim0.01$~Myr fluctuations remain at the level of $\sim15\%$ ($\sim0.06$~dex).

This trend arises because variability on longer timescales is associated with the formation, transport, and accretion of more massive multiphase structures, which produce larger and more coherent accretion events. On shorter timescales, the signal is instead increasingly shaped by smaller, rapidly decorrelating fluctuations in the inner multiphase flow, including cloud interactions and local turbulent stirring, which suppress the variability amplitude.

\subsection{Power spectral density}

\begin{figure}
\centering
\includegraphics[width=0.95\columnwidth]{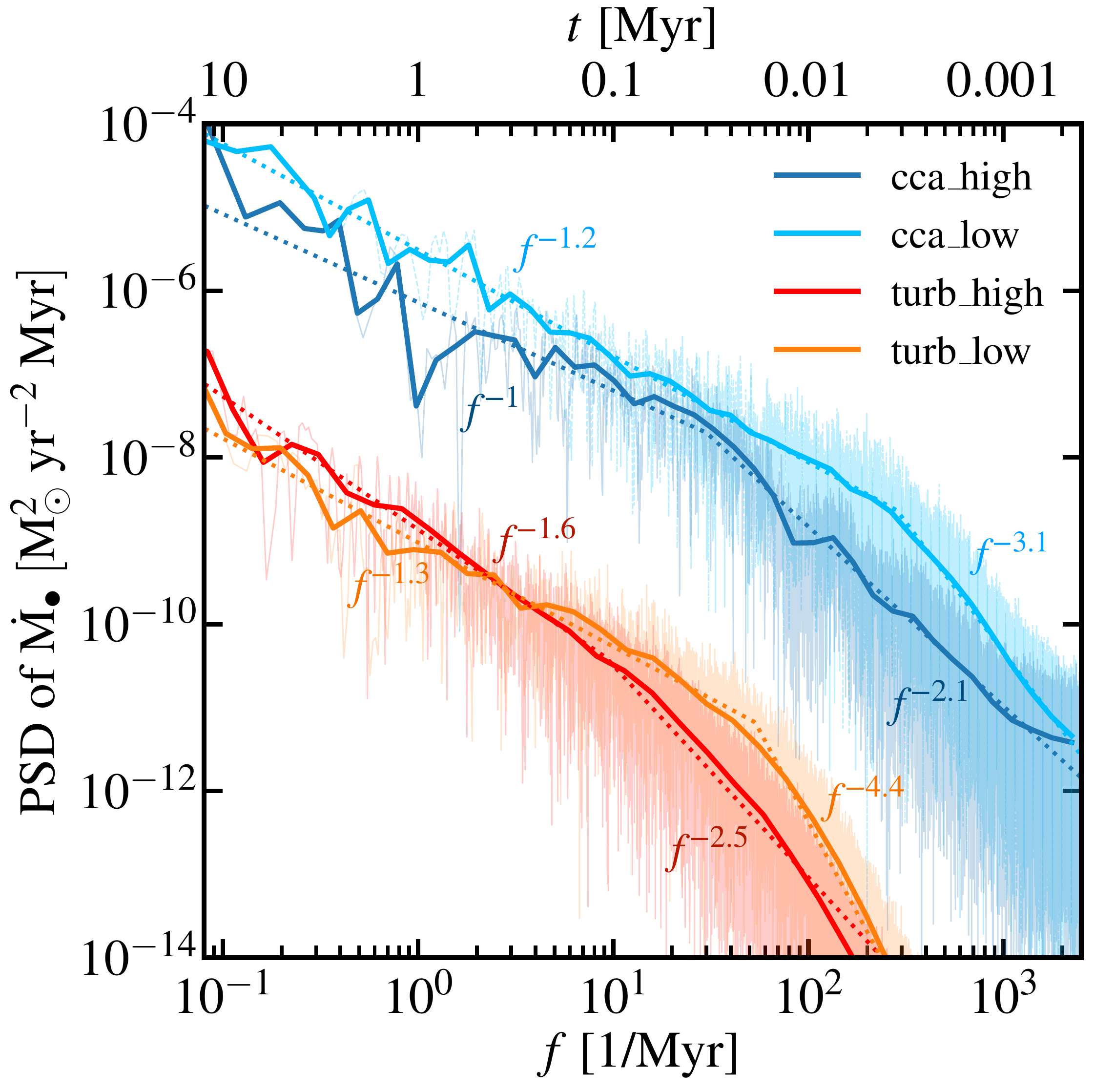} 

\caption{Power spectral density (PSD) of the accretion rate onto the central SMBH for the \lowc\ (light blue) and \highc\ (blue) runs, using the time interval $t\geq t_{\rm rain}$, together with the corresponding turbulence-only cases \turbohigh\ (red) and \turbolow\ (orange). Solid thick lines show logarithmically binned spectra, while thin lines indicate the unbinned periodograms. Dotted lines show the best-fitting broken power-law models over the fitted frequency range. The upper x-axis reports the corresponding timescale $t = 1/f$. In the fiducial simulations (\lowc\ and \highc), the PSD displays pink-noise–like behaviour at low frequencies ($P\propto f^{-1}$), transitioning to red-noise–dominated slopes ($P\propto f^{-\alpha}$ with $\alpha \gtrsim 2$) above the break frequency.}\label{psd}

\end{figure}

A key diagnostic for understanding the variability of the accretion process is the distribution of power in the frequency domain. Figure~\ref{psd} shows the power spectral density (PSD) of the black hole accretion rate $\dot{M_{\bullet}}$ for \lowc\ (light blue), \highc\ (blue), \turbolow\ (orange) and \turbohigh\ (red). The PSD is computed from the squared modulus of the Fourier transform of the accretion rate time series using only times $t\geq t_{\rm rain}$, and quantifies how variability power is distributed across temporal frequencies, $f = 1/t$.

The median PSD of the accretion rate in both fiducial simulations is well described by a broken power law
\begin{equation}
P(f) \propto
\begin{cases}
f^{-\alpha_1}, & f \le f_{\rm b} \\
f^{-\alpha_2}, & f > f_{\rm b}
\end{cases},
\end{equation}
where $\alpha_1$ and $\alpha_2$ are the low- and high-frequency slopes, respectively, and $f_{\rm b}$ is the break frequency.\footnote{Such broken stochastic variability is sometimes discussed phenomenologically in terms of damped-random-walk models in the AGN literature (e.g. \citealt{ZuKochanek2013,IvezicMacLeod2014}).} At low frequencies the spectrum roughly follows $P(f)\propto f^{-1}$, often referred to as pink (or flicker) noise, as previously discovered in CCA simulations by \citet{gaspari2017}, where it was interpreted as the signature of a chaotic, nonlinear, self-similar feeding process. Similar stochastic CCA-like variability has also been suggested for the changing-look AGN Mrk~1018 \citep{DunnMcElroy2025}. This picture is further supported by recent timing studies of AGN obscuration variability. Using long-term X-ray monitoring of the changing-look Seyfert NGC~1365, Serafinelli et al. (in prep.) found that the PSD of the hardness-ratio-variability, which may track obscuration from clumpy circumnuclear medium, follows a similar $P(f)\propto f^{-1}$ scaling during heavily obscured phases.

In the frequency domain, the approximately constant power per logarithmic interval ($fP(f)\approx\mathrm{const}$) indicates that accretion rate variability is distributed comparably across each decade in timescale, with no single temporal scale dominating the fluctuations. This behaviour reinforces the self-similar nature of CCA variability and is also found in the temporal behaviour of turbulent flows in hydrodynamic and magnetohydrodynamic simulations \citep[e.g.][]{herault2015,yuen2025}. At higher frequencies the PSD steepens towards $P(f)\propto f^{-2}$, approaching a red-noise-like regime. In physical terms, long-timescale variability retains the memory of stochastic condensation, cloud interactions, and multiphase inflow, whereas short-timescale fluctuations are progressively decorrelated and damped as structures enter the strongly mixed inner region, where collisions, turbulent stirring, and local inflow dynamics reduce temporal coherence. A negative spectral slope indicates that most of the power resides at low frequencies, corresponding to longer timescales, consistent with the larger amplitude variations of the accretion rate observed over extended periods (Figure~\ref{accretionzoom}). This behaviour is in line with the accretion rate seen in Figure~\ref{accretionlowc}: in the pink-noise regime, variability power increases towards low frequencies but remains distributed across a broad range of timescales, whereas in the red-noise regime the power is increasingly dominated by the longest timescales.
Based on the retrieved PSDs, we can distinguish between three different regimes:

1. Purely turbulent hot mode (sunny weather): in both turbulence-only simulations, the PSD is dominated by red-noise variability. At low frequencies, the spectra show slopes $\alpha_1 \simeq 1.61 \pm 0.14$ (\turbohigh) and $\alpha_1 \simeq 1.35 \pm 0.07$ (\turbolow), steeper than the pink-noise behaviour found in the fiducial runs. At higher frequencies, the PSD further steepens to $\alpha_2 \simeq 2.54 \pm 0.12$ and $\alpha_2 \simeq 4.36 \pm 0.12$, respectively, indicating a strong suppression of short-timescale variability. The two cases differ in break frequency, with \turbohigh\ exhibiting a break at lower frequency ($f_{\rm b}\simeq9\,\mathrm{Myr}^{-1}$) than \turbolow\ ($f_{\rm b}\simeq53\,\mathrm{Myr}^{-1}$), reflecting differences in the strength of the turbulent forcing.

2. Strongly turbulent CCA (\highc, stormy weather): the PSD exhibits a break at $f_{\rm b}\simeq30\,\mathrm{Myr}^{-1}$ (corresponding to a break time $t_b=0.033$ Myr), with low- and high-frequency slopes $\alpha_1\simeq1.06 \pm 0.10$ and $\alpha_2\simeq2.15 \pm 0.14$, respectively. These slopes are very close to the canonical values expected for pink-noise ($\alpha\simeq1$) and red-noise ($\alpha\simeq2$) variability, indicating a transition from long-timescale correlated accretion variability to a strong suppression of power towards short timescales. 
After the break, the PSD slope becomes similar in magnitude to that of the turbulence-only simulations, indicating that on these short timescales the accretion signal becomes increasingly decorrelated and damped in the strongly mixed inner flow.

3. Weakly turbulent CCA (\lowc, rainy weather): the low-frequency PSD remains consistent with pink-noise--like variability, though with a slightly steeper slope ($\alpha_1\simeq1.20 \pm 0.03$). The spectrum exhibits a break at a significantly higher frequency, $f_{\rm b}\simeq284 \,\mathrm{Myr}^{-1}$ ($t_b=0.0035$ Myr), followed by a much steeper high-frequency decay ($\alpha_2\simeq3.11 \pm 0.20$), indicating a strong suppression of short-timescale variability. This mixed behaviour is consistent with the coexistence of long-timescale correlated accretion variability and a more efficient damping of rapid fluctuations in the inner inflow region. Also in this case, the high-frequency variability beyond the break is increasingly decorrelated and damped as the inflow enters the strongly mixed inner region.

The corresponding break times ($t_{\rm b}\simeq 3.5$--$33\,{\rm kyr}$) are consistent with an inner-flow origin. For our $M_\bullet \sim 2.8\times10^8\,{\rm M}_\odot$ SMBH, kyr-scale variability matches characteristic dynamical/crossing times on parsec scales,
\begin{equation}
t_{\rm dyn} \equiv 2\pi \left(\frac{r^3}{G M_\bullet}\right)^{1/2}
\simeq 5.6\,{\rm kyr}\,
\left(\frac{r}{1\,{\rm pc}}\right)^{3/2}
\left(\frac{M_\bullet}{2.8\times10^8\,{\rm M}_\odot}\right)^{-1/2},
\end{equation}
implying that the measured $t_{\rm b}$ corresponds to characteristic radii of $\sim$\,$0.7$--$3$ pc (or to comparable free-fall/crossing scales at slightly larger radii, depending on geometry and thermodynamics).
In this view, $f_{\rm b}$ identifies the timescale below which the accretion signal progressively loses temporal coherence as multiphase structures enter the strongly mixed, collision-dominated inner region; rapid fluctuations are then more efficiently decorrelated and damped, steepening the PSD at high frequencies. At the same time, the absolute value of $t_{\rm b}$ should not be over-interpreted, since it lies only modestly above the sink scale in temporal terms; the more robust result is the relative shift in $f_{\rm b}$ between \lowc\ and \highc. Crucially, $t_{\rm b}$ is far longer than the Courant timestep and sink-update cadence, yet far shorter than the driving correlation time ($t_{\rm corr}=30\,{\rm Myr}$) and the large-scale eddy turnover time, indicating that the break is unlikely to be set by numerical update frequencies or by the outer forcing timescale. The systematic shift in $f_{\rm b}$ between \lowc\ and \highc\ further supports a physical sensitivity to the turbulence regime and multiphase coupling.

This low-frequency behaviour is in agreement with the CCA variability reported by \citet{gaspari2017} ($\mathcal{M}\sim0.4$ weather), who found that the accretion rate power spectral density follows a nearly scale-free pink-noise law, with a fitted slope of $-1.09$, and interpreted it as the signature of a chaotic, nonlinear, self-similar process. Our results corroborate that picture: the low-frequency slopes of the two fiducial runs ($\alpha_1\simeq1.06$ in \highc\ and $\alpha_1\simeq1.20$ in \lowc) remain consistent with that CCA discovery. At the same time, the present higher-resolution simulations extend that result by showing that the PSD may be more accurately described as a broken power law, with a turbulence-dependent break frequency and a steeper high-frequency tail. In this sense, \citet{gaspari2017} established the pink-noise backbone of CCA variability, while the present work resolves how the transition to short-timescale damping depends on the strength of turbulence and on the degree of inner multiphase coupling.

Taken together, the different break frequencies mark the transition from correlated variability to a more rapidly decorrelating regime. In \lowc, the extension of the pink-noise spectrum to higher frequencies indicates that temporal correlations persist to shorter timescales, consistent with a more coherent inner inflow. In \highc, the earlier transition to red-noise behaviour implies that correlations are confined to longer timescales, with short-timescale variability more efficiently suppressed by stronger turbulent mixing and damping.
Future longer-duration and higher-resolution {\sc BlackHoleWeather} runs will help test whether sustained precipitation shifts $f_{\rm b}$ upward, prolonging the pink-noise regime as micro-scale variability strengthens.

\section{Kinematics -- CCA diagnostics}\label{cca_diag}

We now turn to the kinematic diagnostics of multiphase condensation. To connect the thermodynamic state of the halo to the dynamical state of the inflow, we use two complementary tools: the $\mathcal{C}$-ratio, which compares radiative cooling and turbulent eddy timescales, and the kinematic phase-space diagram (k-plot), which traces how the different thermal phases populate line-shift/line-broadening space. Together, these diagnostics provide an observation-ready framework to identify where condensation is triggered and how the resulting multiphase structures couple to turbulent motions.

\subsection{$\mathcal{C}$-ratio}\label{sec:cratio_profiles}

Understanding where and when multiphase condensation occurs in hot gaseous halos is central to connecting simulations of AGN feeding and feedback with observations in different bands. As discussed in B26a, the emergence of cold filaments and clouds is closely linked to the interplay between radiative cooling and turbulent mixing, rather than gravity alone. To quantify this interplay, the $\mathcal{C}$-ratio was introduced \citep{gaspari2018} and defined as

\begin{equation}
\mathcal{C} \equiv \frac{t_{\text{cool}}}{t_{\text{eddy}}}.
\end{equation}
$t_{\text{cool}}$ is the plasma cooling time, defined as

\begin{equation}
t_{\text{cool}} = \frac{3k_{\mathrm{B}}T}{n_e \Lambda},
\end{equation}
where $k_{\mathrm{B}} = 1.38 \times 10^{-16}\,\mathrm{erg\,K^{-1}}$ is the Boltzmann constant, $T$ is the gas temperature, $n_e$ is the electron number density and $\Lambda$ is the cooling function described in Section \ref{num}.
The eddy \textit{gyration} time $t_{\text{eddy}}$ is the characteristic time required for a (subsonic/solenoidal) turbulent eddy to complete one turnover, i.e. to circulate once over its own size. It therefore quantifies how rapidly subsonic turbulence cascades stirring significant perturbations on a given spatial scale \citep{gaspari2018}:
\begin{equation}
t_{\text{eddy}}
 = 2\pi \frac{r^{2/3}L^{1/3}}{\sigma_{v,L}},
\end{equation}
where $L$ is defined as the turbulence injection scale and $\sigma_{v,L}$ is the velocity dispersion at the injection scale. 
The formula is derived under the assumption that turbulent velocities follow a \citet{kolmogorov1941} cascade, i.e. $\sigma_v(l) \propto l^{1/3}$, as confirmed in simulations of subsonic turbulence in hot halos
\citep[e.g.][]{gasparichurazov2013}, and that at a given radius within a galaxy or cluster atmosphere, the characteristic eddy size is expected to be of the order of the local radius, since turbulent structures larger than $r$ cannot be self-contained within that region, hence $l\sim r$. 
For reference, at the driving scale adopted in these simulations, $L_{\rm inj}\simeq 12.5$ kpc, the corresponding \textit{outer}-eddy gyration times in the hot, hard X-ray phase are of order
$t_{\rm eddy}(L_{\rm inj}) \sim 0.3$ Gyr in \highc\ and $\sim 0.9$ Gyr in \lowc.

\begin{figure*}
\centering
\includegraphics[width=0.94\textwidth]{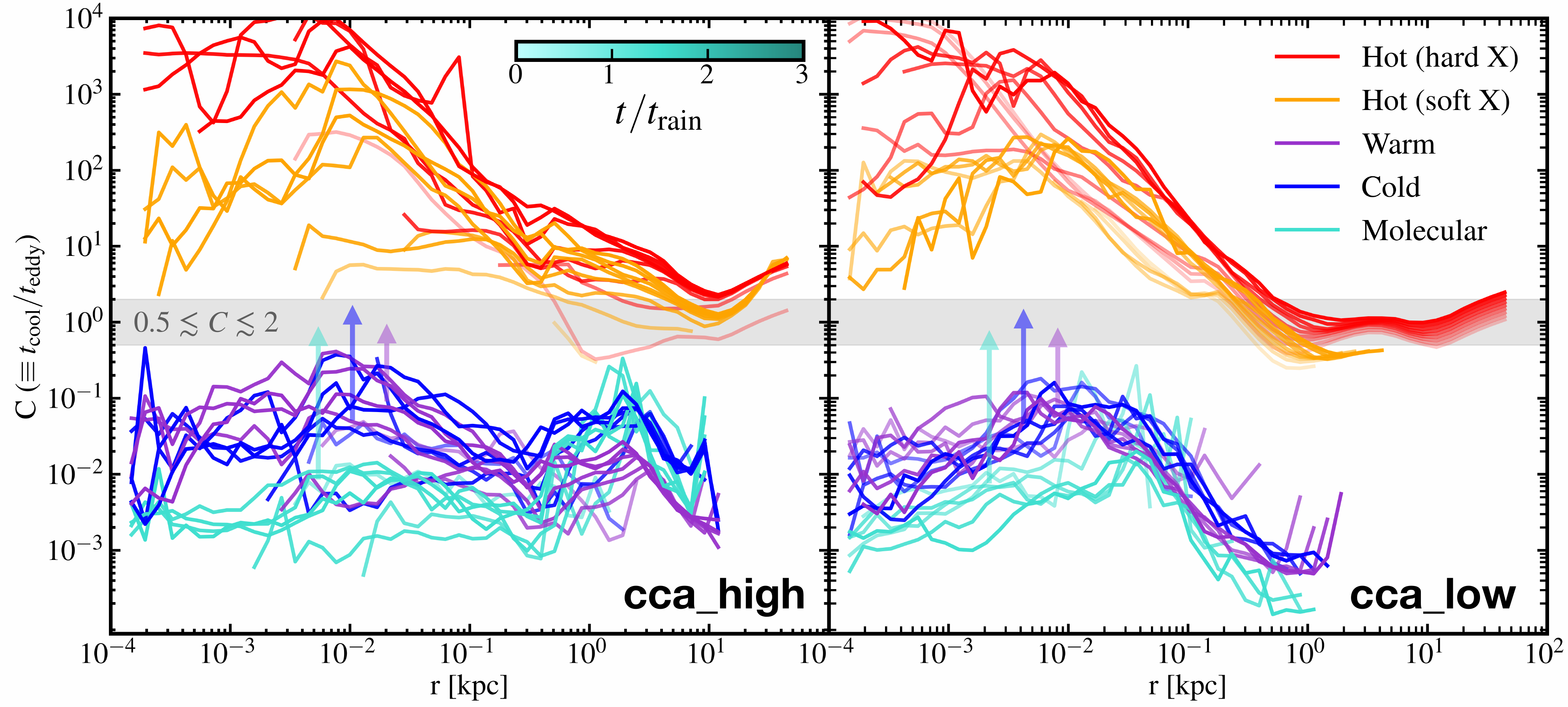} 

\caption{Radial profiles of the $\mathcal{C}$-ratio ($\equiv t_{\mathrm{cool}}/t_{\mathrm{eddy}}$) in the \highc\ (left) and \lowc\ (right) simulations, shown for different gas phases and colour–coded by observational band (see Table \ref{tab:thermal_phases}). For each phase, the sequence of shades traces the temporal evolution from the start of the run to $t/t_{\mathrm{rain}} = 3$, as indicated by the colour bar at the top. The horizontal grey band marks the range $0.5 \lesssim \mathcal{C} \lesssim 2$ (i.e.~a reasonable 0.3\,dex scatter on unity threshold), where multiphase condensation is expected to be most efficient.}\label{cratio}

\end{figure*}

Unlike traditional criteria such as $t_{\rm cool}/t_{\rm ff}$ \citep{mccourt2012,gaspari2012,sharma2012,voit2015}, where $t_{\rm ff}$ is the free-fall time and depends on uncertain gravitational potentials, the $\mathcal{C}$-ratio offers a more directly turbulence-based diagnostic. Observationally, both ingredients of the $\mathcal{C}$-ratio can in principle be constrained with multi-wavelength spectroscopy: X-ray data can provide cooling-time profiles $t_{\rm cool}(r)$, while turbulent velocity dispersions can be inferred from line broadening or resolved gas kinematics using facilities such as XRISM/Resolve, MUSE, JWST, and in the future NewAthena/X-IFU. The $\mathcal{C}$-ratio therefore offers a powerful, testable way to identify regions susceptible to multiphase condensation in the intracluster, intragroup, and circumgalactic medium.

When $\mathcal{C} \ll 1$ (i.e. $t_{\rm cool} \ll t_{\rm eddy}$), radiative cooling is much faster than turbulent stirring, so the gas cools coherently before turbulence can act. Rather than forming a turbulence-driven multiphase medium, the system approaches a classical cooling-flow--like regime, characterised by rapid, large-scale inflow of cooling gas. In contrast, when $\mathcal{C} \gg 1$ (i.e. $t_{\rm cool} \gg t_{\rm eddy}$), turbulent mixing operates faster than cooling, erasing density perturbations before they can cool, thereby suppressing multiphase condensation and maintaining the gas in a hot, single-phase state. When $\mathcal{C} \equiv t_{\rm cool}/t_{\rm eddy} \approx 1$, i.e. when the cooling time is comparable to the eddy turnover time (usually within a $\sim$0.3\,dex scatter from unity; \citealt{gaspari2018}), turbulent density perturbations have enough time to grow and condense before being mixed or reheated. In this regime, multiphase condensation develops, producing extended warm and cold filaments, as observed in a wide range of galaxies, groups, and clusters \citep[e.g.][]{maccagni2021,juranova2020,temi2022,olivaressalome2022,singha2023,Mehdipour2023,lepore2025}.

Figure \ref{cratio} shows the radial profiles of the $\mathcal{C}$-ratio for the \highc\ (left panel) and \lowc \ (right panel) simulations across multiple gas phases, colour-coded by temperature: hot (hard and soft X-ray), warm, cold, and molecular phases, as described in Table \ref{tab:thermal_phases}. Each curve corresponds to a different time from the start of the simulation to $t/t_{\text{rain}}=3$, with the spread within each colour indicating the passage of time.
It is important to note that, in this analysis, all plotted $\mathcal{C}$-ratios use the hot-phase eddy time, $t_{\rm eddy,hot}$, as a common normalisation. This is appropriate for tracing coupling to the large-scale/\textit{ensemble} turbulent cascade, but it likely underestimates the \textit{intrinsic} $\mathcal{C}$-ratio of the colder phases, which reside in progressively smaller filaments and molecular clouds. $t_{\rm cool}$ for each phase is computed as a mass-weighted average within spherical radial shells, while the hot-phase velocity dispersion $\sigma_{v,\rm hot}(r)$ is evaluated in the same shells after subtracting the bulk motion.

We begin by examining the hot phase. In the \highc\ simulation (left panel), the $\mathcal{C}$-ratio is approximately flat at large radii ($r \gtrsim 10$ kpc), with values of $\sim 1$--$2$. Moving inward, however, it rises steeply and reaches $\sim 10^3$--$10^4$ in the central region. This increase is primarily driven by the rapid inward decline of the eddy-turnover time, since smaller eddies stir the gas on progressively shorter timescales ($t_{\rm eddy}\propto l^{2/3}$), more rapidly than the cooling time decreases as the density rises. As a result, the hard X-ray phase remains turbulence-dominated over most radii, with $\mathcal{C}\gtrsim 1$, implying that radiative cooling is generally inefficient unless perturbations reach nonlinear amplitudes.

This condition is met when the hard X-ray gas enters the condensation band, $\mathcal{C}\approx 0.5$--$2$, which in \highc\ occurs mainly at $r \gtrsim 1~\mathrm{kpc}$, especially at earlier times ($t<t_{\rm rain}$). In this regime, nonlinear cooling can efficiently develop and drive the transition towards the soft X-ray phase ($1.16\times10^6 \leq T < 5.8\times10^6$ K). Physically, the densest hard X-ray regions cool first into the soft X-ray regime, while the residual hard X-ray gas is left at lower density and therefore becomes progressively less prone to further condensation. Consistently, the hard X-ray $\mathcal{C}$-ratio increases with time, indicating that nonlinear perturbations become less effective in this phase.

The soft X-ray gas, which forms directly out of the cooling hard X-ray phase, follows a similar radial trend but with systematically lower $\mathcal{C}$ values, reaching $\mathcal{C}\approx 10$--$100$ in the central core. Most importantly, it remains close to the condensation band, with $0.5 \lesssim \mathcal{C} \lesssim 2$ over much of the radial range $\sim 5$--$15~\mathrm{kpc}$, where it acts as the immediate thermodynamic precursor of further cooling towards the warm and cold phases. This highlights the soft X-ray component as the key transition layer of the condensation cascade and the gateway to filamentary multiphase structure, in agreement with recent observational results \citep{olivares2025}. Figure~\ref{cratio} therefore suggests a self-regulating thermodynamic sequence: condensation selectively removes the low-$\mathcal{C}$ tail of the hard X-ray distribution, while the remaining hot atmosphere shifts to larger $\mathcal{C}$ values and becomes increasingly mixing-dominated.

In the \lowc\ simulation (right panel), the hard X-ray phase maintains a nearly constant $\mathcal{C}$-ratio of $\approx 0.5$--$1$ over most radii ($r \gtrsim 0.1$ kpc), before rising steeply to $\approx 10^3$--$10^4$ in the central region. Overall, these values are systematically lower than in \highc, implying that nonlinear cooling can develop more easily. At early times, the resulting condensation and soft X-ray formation are largely restricted to within $\sim 10$ kpc. The soft X-ray phase itself begins at $\mathcal{C}\approx 0.4$--$1$ at larger radii, then rises sharply towards the centre up to $\mathcal{C}\sim 100$. This indicates that in \lowc\ the soft X-ray component can still enter the condensation band, but over a narrower radial range and with a more centrally concentrated evolution than in \highc.

Taken together, the $\mathcal{C}$-profiles identify the soft X-ray phase as the key thermodynamic gateway of condensation. In \highc, this phase remains within the canonical condensation band over a broad radial range, enabling sustained formation of warm filaments and colder structures out to several kpc. In \lowc, the same transition is more centrally concentrated and develops later, consistent with a weaker and more spatially confined condensation cascade. The $\mathcal{C}$-ratio therefore captures the main difference between the stormy and rainy regimes: not whether cooling occurs, but how broadly turbulence and cooling remain matched across the halo. A complete assessment, however, requires combining thermodynamic diagnostics such as the $\mathcal{C}$-ratio with kinematic and temporal probes, including
k-plots, multiphase inflow measurements, and accretion rate variability/PSD analyses.

In contrast, the warm and cold phases exhibit $\mathcal{C}$-ratios several orders of magnitude below unity, typically in the range $10^{-2}$–$10^{-1}$ across most radii in both simulations (given the substantially shorter cooling times but the same macro-scale eddy time). Notably, the warm and cold neutral phases differ substantially from the molecular phase. The warm and cold components display nearly identical $\mathcal{C}$-values—systematically higher than those of the radio-emitting gas—typically around $10^{-2}$–$10^{-1}$. This intermediate regime bridges the transition between the cooling and condensed states, suggesting that these phases act as precursors from which molecular gas originates.
Their $\mathcal{C}$-ratio profiles show a mild central rise, peaking at $r \sim 10^{-2}$–$10^{-1}$ kpc before flattening outwards, implying that cold neutral gas remains partially coupled to the turbulent cascade within the inner core. This intermediate state reflects a dynamically mixed population of recently cooled structures and marginally unstable gas still exchanging energy and momentum with the hot background. The molecular phase instead shows lower $\mathcal{C}$-ratios, around $10^{-3}$–$10^{-2}$ in the central region, while overlapping with the warm and cold phases between 1 and 10 kpc. In the \lowc\ simulation, the three phases display comparable $\mathcal{C}$-ratios, though the gas extends only to $\sim 1$ kpc. 

The consistently low $\mathcal{C}$-values of the molecular phase mark it as the endpoint of the instability cascade, namely dense condensates that are increasingly decoupled from the large-scale turbulent flow.
Radially, its $\mathcal{C}$-profile remains flat or mildly declining, consistent with its confinement to the inner cooling region. 
If the characteristic size decreases by a factor $\ell/\ell_{\rm hot}\sim 10^{-2}$--$10^{-3}$ from the hot outer scale to cold/molecular structures, then the effective turnover time is shorter by a factor $t_{\rm eddy}(\ell)/t_{\rm eddy,hot}\sim (\ell/\ell_{\rm hot})^{2/3}\sim 10^{-1}$--$10^{-2}$. This would increase the phase-dependent $\mathcal{C}$-ratio by $\sim 10$--$100$, potentially bringing the cold/molecular gas from the plotted $\mathcal{C}\sim 10^{-3}$--$10^{-2}$ up to $\mathcal{C}\sim 0.1$--$1$ near the characteristic condensation region (see coloured arrows in Figure~\ref{cratio}). A more quantitative assessment requires measuring the typical structure size $\ell$ for each phase and their internal velocity dispersion, which we defer to future work.

Overall, for direct comparison with observational applications of the $\mathcal{C}$-ratio, the most relevant quantities are therefore the hard- and especially soft-X-ray profiles, since they trace the ambient hot atmosphere in which condensation is triggered. By contrast, the plotted $\mathcal{C}$-values of the warm, cold, and molecular phases should be regarded as lower limits when a common hot-phase eddy time is adopted, because these phases reside in progressively smaller structures with intrinsically shorter turnover times.

\begin{figure*}
\centering
\includegraphics[width=\textwidth]{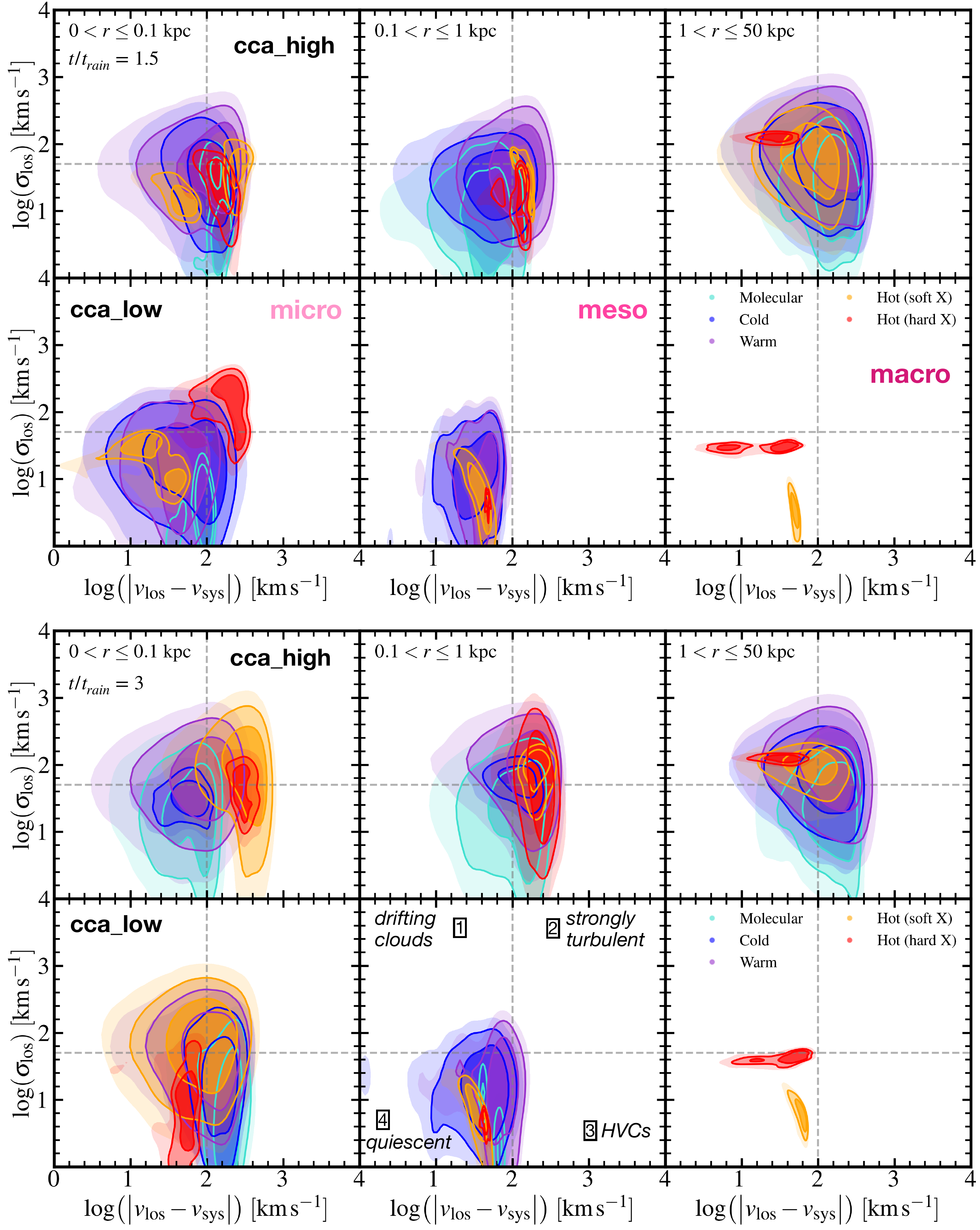} 

    \caption{Multi-scale k–plots of the gas in \highc\ and \lowc\ computed at $t/t_{\text{rain}}=1.5$ (top) and at $t/t_{\text{rain}}=3$ (bottom)  in three radial scales. Each panel shows $\log(|v_{\mathrm{los}}-v_{\mathrm{sys}}|)$ versus $\log(\sigma_{\mathrm{los}})$. Coloured shaded regions and contours give the 2D distribution of gas in each thermal phase (colour–coded as in the legend); increasing opacity and darker contours mark the 85th, 92nd, and 97th percentiles of the underlying histogram, respectively. Gas in the lower–left corner (low velocity offset and dispersion) traces quiescent gas, whereas gas towards the upper–right corresponds to high–velocity, high–dispersion motions characteristic of turbulent, collisionally mixed CCA.}\label{kplot}

\end{figure*}

\subsection{Kinematic plot (k-plot)}
To characterise the velocity structure of the multiphase medium, we use kinematic diagnostics known as k–plots \citep{gaspari2018,maccagni2021}, which display the line-of-sight velocity offset from the systemic velocity (here set to zero), $|v_{\mathrm{los}}-v_{\mathrm{sys}}|$, against the corresponding line-of-sight velocity dispersion $\sigma_{\mathrm{los}}$. 
The k-plots are computed by using mass-weighted values for each phase and radial range, projecting along the z-axis.

Figure~\ref{kplot} presents k–plots at two representative epochs, $t/t_{\rm rain}=1.5$ (top panels) and $t/t_{\rm rain}=3$ (bottom panels), for the three characteristic spatial scales: micro ($r\leq0.1$ kpc), meso ($0.1<r\leq1$ kpc), and macro ($1<r\leq50$ kpc), in the two turbulence regimes, \highc\ and \lowc. Gas phases are colour-coded by temperature (see Table \ref{tab:thermal_phases}). The dashed lines divide the plane at $|v_{\mathrm{los}}-v_{\mathrm{sys}}|=100~\mathrm{km \ s^{-1}}$ and $\sigma_{\mathrm{los}}=50~\mathrm{km \ s^{-1}}$, separating distinct kinematic regimes. Within the framework of CCA, k–plots provide a compact diagnostic of the dynamical state of the inflow (or outflow). Quiescent gas (no chaotic motions) occupies the lower-left region of kinematic space, where both velocity offsets and dispersions are small, while strongly turbulent flows populate the upper-right, where both quantities are large. The morphology, extent, and overlap of the phase-dependent loci in k–space therefore encode the transition from a quiet state to chaotic accretion, with developed CCA rain typically occupying the central $\sim$50–100 km s$^{-1}$ locus \citep[cf.][]{gaspari2018}.
The remaining quadrants trace mixed kinematic states: gas with relatively large dispersion but modest centroid shift is consistent with turbulence-dominated, ensemble-like motions, whereas gas with large centroid shift but narrower lines is more naturally associated with discrete, rapidly infalling cloud components, analogous to the high-velocity-cloud (HVC) population discussed in absorption studies \citep[e.g.][]{tremblay2016,tremblay2018}.

On the micro-scale ($r \leq 0.1$ kpc; first column), the \highc\ simulation exhibits intermediate velocity offsets and significant velocity dispersions across all gas phases, consistent with a strongly perturbed central environment.
The hot gas transitions from the centre to the high velocity offset region of the k–plot at $t/t_{\rm rain}=3$, with typical centroid values of order $|v_{\rm los}-v_{\rm sys}|\approx 200$–250 km s$^{-1}$, both with low and high velocity dispersions. On the other hand, warm, cold and molecular gas have substantially lower offsets, $|v_{\rm los}-v_{\rm sys}|\sim 30$–100 km s$^{-1}$, with the molecular gas moving towards the quiescent quadrant over time. At both times, cold and molecular phases have the centroids roughly coincident, showing a strong correlation in the velocity space. Despite this separation in bulk motion between colder and hotter phases, the line-of-sight velocity dispersions remain comparable. In particular, the different phases show overlapping $\sigma_{\rm los}\sim 20$--60 km s$^{-1}$, indicating a strong multiphase kinematic coupling.
An important nuance of the micro-scale kinematics is that the phases can remain strongly coupled in dispersion while being partially separated in centroid velocity. This indicates that, near the centre, the multiphase gas still shares a common turbulent environment even when individual clouds or filaments develop distinct bulk motions relative to the hotter background.
In contrast, the \lowc\ simulation shows a much more ordered micro-scale kinematic structure (given the lower driven turbulence variance). All condensed phases collapse into compact loci around $|v_{\rm los}-v_{\rm sys}|\lesssim 50$–100 km s$^{-1}$ and $\sigma_{\rm los}\lesssim 20$–30 km s$^{-1}$. In this regime, cold and molecular gas can still reach relatively large bulk velocities, but their internal dispersions remain very small.

On meso-scales ($0.1<r\leq1$ kpc; second column), the kinematic differences between the two simulations become more pronounced. In \highc, all gas phases occupy broad and strongly overlapping regions, indicating efficient multiphase coupling also at these scales. From $t/t_{\rm rain}=1.5$ to $t/t_{\rm rain}=3$ all the phases transition from low $|v_{\rm los}-v_{\rm sys}|$ and $\sigma_{\rm los}$ to higher values moving diagonally from the quiescent to the strongly turbulent quadrant. This diagonal migration in the kinematic plane reflects the progressive increase in both bulk velocities and internal turbulence driven by the stronger stirring at meso-scales. It is precisely at these radii that condensation becomes most efficient, leading to enhanced multiphase fragmentation and the formation of the extended cold structures characteristic of the stormy regime. The hot gas spans large velocity offsets and dispersions, with typical values of $|v_{\rm los}-v_{\rm sys}|\approx 100$–300 km s$^{-1}$ and $\sigma_{\rm los}\sim 80$–200 km s$^{-1}$. Intermediate-temperature gas fills the central part of the diagram, continuously connecting the hot and cold components, as this warm gas is typically formed at the edge of filaments and clumps.
In contrast, the meso-scale kinematics of the \lowc\ simulation remain in the more quiescent quadrant, showing feeble evolution over time. All phases collapse into narrow loci with relatively small velocity dispersions, typically $\sigma_{\rm los}\lesssim 10$ km s$^{-1}$, and limited velocity offsets. 
The small velocity offsets and dispersions, reflecting the weaker stirring, explain why meso-scale condensation remains weak and spatially confined in \lowc. In contrast to \highc, the phase loci remain compressed in the quiescent region and do not develop the broad multiphase spread associated with strongly nonlinear cooling at these radii. As a result, cold gas remains largely confined to the central region, yielding a compact, rainy configuration.

Looking at macro-scales ($1<r\leq50$ kpc; third column), both simulations show close to no evolution over time. In the \highc\ simulation, the multiphase medium continues to occupy broad and partially overlapping regions in the k-plot (now becoming a more ensemble description). The hot gas spans moderate-to-large dispersions, with typical values $\sigma_{\rm los}\approx 100$–120 km s$^{-1}$, while the soft X-ray and intermediate-temperature gas retain wide spreads in velocity offset. Warm and cold components develop extended tails towards higher dispersions, indicating that large-scale turbulent motions still perturb the inflowing material and maintain multiphase coupling well beyond the central region. At this scale we can see very well how the k--plot directly traces the condensation cascade, with gas progressively cooling out of the hot, high-dispersion phase and transitioning towards colder components characterised by systematically lower velocity dispersions (as expected during a top-down turbulence cascade), down to the molecular phase.
In contrast, the macro-scale kinematics of the \lowc\ simulation are again more compact and only the weakly stirred hot phase is present, which clusters into a narrow locus with relatively small dispersions, $\sigma_{\rm los}\lesssim 40$ km s$^{-1}$, and limited spreads in velocity offset, especially in the soft X-ray phase.

Taken together, the k-plots across micro, meso, and macro-scales show that both simulations sustain CCA-like multiphase inflow in the central region, but only \highc\ maintains strong multiphase coupling and extended condensation out to macro-scales. Strong (still subsonic $\mathcal{M}\sim0.4$) turbulence in \highc\ produces broad, overlapping kinematic loci and preserves a dynamically connected condensation cascade across radii, whereas in \lowc\ ($\mathcal{M}\sim0.15$) the multiphase gas remains largely confined to the inner halo and the outer regions are dominated by a comparatively quiescent hot phase. The k-plot therefore provides a direct kinematic discriminator between spatially extended, stormy CCA and centrally concentrated, rainy accretion.

A useful comparison can be made with recent {\sc BlackHoleWeather} simulations including explicit AGN feedback (C26b). In both cases, the stormy weather is associated with broad and overlapping multiphase loci in the k-plane, whereas the rainy weather remains more compact and kinematically coherent. In C26b, jet-driven uplift and recycling further broaden the phase distributions and enhance the stochastic wandering of the condensed gas across the kinematic quadrants, particularly in the inner kpc. This suggests that large-scale turbulent stirring alone is sufficient to establish the main kinematic signatures of multiphase CCA, while explicit jet feedback primarily amplifies the circulation, recycling, and phase mixing of the condensates.

These two main CCA diagnostics, $\mathcal{C}$-ratio and k-plot, thus provide a compact and observation-ready framework to connect where condensation is triggered to how the resulting multiphase gas moves across scales.
Among all radii, the meso-scale emerges as the clearest dynamical separator between the two halo-weather states: it is here that strong turbulence sustains broad, overlapping multiphase kinematics and extended condensation, whereas weak turbulence leaves the flow confined to a compact, more quiescent locus.

\section{Comparison with observations}\label{sec:synthesis}
A growing body of multi-wavelength observations supports the emerging picture of CCA in group- and cluster-scale atmospheres (e.g. Perseus, M~87, Centaurus, Abell~1795, Abell~2597, Hydra-A, PKS~0745$-$191). In cool-core clusters, deep Chandra and optical imaging reveal extended, filamentary multiphase structures, with warm H$\alpha$ nebulae closely aligned with soft X-ray filaments and molecular gas over projected scales of several to tens of kpc \citep[e.g.][]{olivaressalome2022,olivares2025}. The tight spatial correspondence between hot, warm, and cold phases is naturally explained if the filaments form in situ from the hot atmosphere through a condensation and mixing cascade, rather than being purely externally supplied. These observational trends are consistent with the stormy CCA picture, where turbulence maintains multiphase coupling and fragmented transport over a wide radial range.

At lower halo masses, ALMA observations of group-centre ellipticals provide a complementary view. \citet{TemiAmblard2018} detect numerous CO-emitting clouds in NGC~5044, NGC~5846, and NGC~4636. The CO structures are highly fragmented on scales of $\sim 10^2$~pc and have virial parameters $\alpha_{\rm vir}\gg 1$, indicating that they are not classical bound giant molecular clouds but unbound molecular associations embedded in a turbulent medium. The close spatial correspondence between CO, H$\alpha$+[N,\textsc{ii}], and soft X-ray emission strongly supports in-situ condensation from the hot halo and a CCA-like origin of the multiphase gas. The total molecular mass that we find in \highc, $M_{\mathrm{H}2}\simeq 5.5\times10^{7}$ M$_{\odot}$, lies at the upper end of the observed group range, as expected for a strongly cooling atmosphere and sustained multiphase precipitation.

More direct evidence for CCA-like `rain' comes from molecular absorption measurements against bright nuclear continuum sources. In the Abell~2597 BCG, ALMA observations reveal cold, clumpy clouds seen in absorption and moving inward at a few $\times 10^{2}\,{\rm km\,s^{-1}}$ on sub-kpc to $\lesssim$100 pc scales \citep{tremblay2016,tremblay2018}. A larger ALMA absorption survey of cool-core BCGs finds multiple narrow absorption components with velocities biased towards inflow, consistent with small, discrete clouds feeding the nucleus in a chaotic manner \citep{rose2019}. These observations map naturally onto the ``HVC'' and central CCA loci in the k-plot found in this work (Figure \ref{kplot}), supporting the interpretation that at least part of the fueling proceeds through compact, rapidly varying cold structures rather than a smooth hot inflow.

A recent observational study of multiphase gas in group-centred galaxies \citep{temi2026} resolves distinct dust and gas structures within the central few kpc, revealing a dynamically unsettled medium. Chaotically distributed dusty fragments and filaments are present, consistent with cold gas in a transient orbital state at several kpc radii, where the free-fall time is $\sim 10^7$ years. Off-centre orbiting molecular clouds, with radial velocities matching nearby H$\alpha$ emission, indicate a true physical link between dust, molecular, and warm ionised gas, disfavouring axisymmetric inflow in favour of chaotic, clumpy condensation in the hot X-ray halo. The presented simulations are broadly consistent with these observations of group-central galaxies, with the diversity of thermodynamic and kinematic properties among individual systems naturally emerging within the explored parameter range.

The $\mathcal{C}$-ratio diagnostic has been applied to a variety of real systems hosting multiphase gas. Using X-ray and optical observations of galaxy clusters, \citet{OlivaresSalome2019} derived $\mathcal{C}$-profiles with typical values $\mathcal{C}\approx0.3$–$1.7$, while for galaxy groups \citet{olivaressalome2022} found $\mathcal{C}\approx0.5$–$1.6$ at $r\sim10$ kpc. Similar values are reported for rotating early-type galaxies: \citet{juranova2020} showed that systems hosting cold gas discs exhibit $\mathcal{C}\sim0.5$–$2$, with values close to unity in regions where cold gas is observed, even when classical $t_{\rm cool}/t_{\rm ff}$ criteria are not satisfied. Extending this picture, \citet{temi2022} found that galaxies with multiphase gas are characterised by $\mathcal{C}$-ratios of order unity within the inner $\sim$1–10 kpc, whereas galaxies devoid of cold gas typically show significantly larger values, indicative of overheated atmospheres; in the same sample, NGC~4406 reaches $\mathcal{C}\sim10$, suggesting that precipitation is very weak and that the cool gas may instead have an external origin. A mildly super-unity case was reported by \citet{ubertosi2023}, who measured $\mathcal{C}\sim1.5$ at $r=10$ kpc in a cluster hosting H$\alpha$ filaments. Another super-unity but more ambiguous case is the multiphase orphan cloud in Abell~1367, for which \citet{GeLuo2021} found $\mathcal{C}=2.3$--$4.5$; however, in that system the cool gas may be related to stripped material and mixing, rather than to in-situ condensation. A direct measurement of the condensation ratio was recently obtained for the cool-core cluster Abell~2667 by \citet{lepore2025}, who found $\mathcal{C}\simeq0.4$ within the central $\sim25$ kpc, consistent with a system approaching or crossing the condensation threshold. Finally, consistent results were obtained for the nearby group-central galaxy Fornax~A, where \citet{maccagni2021} measured $\mathcal{C}\approx1$ in regions coincident with ionised and molecular filaments. Overall, these observational studies consistently indicate that multiphase condensation preferentially occurs in regions where turbulent and cooling timescales are comparable, in broad agreement with the $\mathcal{C}$-ratio values measured in our simulations.

Complementing the $\mathcal{C}$-ratio constraints, spatially resolved observations probe the kinematics of multiphase gas and enable a direct comparison to k-plots. IFU and ALMA studies of cool-core BCGs and massive ellipticals commonly find line-of-sight dispersions $\sigma_v \sim 50$--200 km s$^{-1}$ and bulk velocities of a few $\times10^{2}$ km s$^{-1}$ on kpc scales, often with multiphase components that are co-spatial and share correlated kinematics \citep[e.g.][]{rose2019,russell2019,Vantyghem2019,maccagni2021,olivaressalome2022,temi2022,singha2023}, consistent with a prevalent CCA rain. 
In this observational context, the broad, overlapping cold and warm loci in \highc\ mirror the tangled, filament-dominated kinematics seen in strongly disturbed cool cores, where injected turbulence (typically via AGN jets/outflows; e.g. \citealt{Mehdipour2023}) promote widespread multiphase coupling. Conversely, the narrower, more stratified loci in \lowc\ resemble systems in which cold gas forms a more ordered structure (often rotation-dominated) within a calmer hot halo, as suggested in rotating early-type galaxies with disc-like multiphase reservoirs \citep[e.g.][]{juranova2019,juranova2020}. In this way, our sub-parsec CCA simulations provide a unified physical framework to interpret the observed diversity of multiphase morphologies and kinematics in galactic nuclei.

In the hot phase, XRISM/Resolve is now directly measuring the velocity dispersions and bulk flows that enter turbulence-regulated condensation models. In relaxed cool cores such as Abell~2029, Resolve finds a subsonic ICM with $\sigma_v\simeq169$ km s$^{-1}$ and $|v_{\rm bulk}|<100$ km s$^{-1}$ \citep{xrism2025}, while in the archetypal radio-mode system Hydra-A it measures $\sigma_v\simeq164$ km s$^{-1}$ across most of the cooling volume \citep{RoseHydraA_2025}. In Centaurus, Resolve detects a structured core bulk flow (sloshing-driven ``wind'') of $\sim$130--310 km s$^{-1}$ within $\sim$30 kpc, yet with low small-scale dispersion ($\sigma_v\lesssim120$ km s$^{-1}$) \citep{XRISM_Centaurus_2025}, highlighting that coherent motions can be substantial even when turbulent pressure support remains modest. 
These values lie in the same $\sigma_v\sim 50$--200 km s$^{-1}$ band that characterises the hot phase in our simulations on kpc scales, supporting a CCA picture where multiphase precipitation and intermittent clumpy feeding occur without requiring supersonic stirring, consistent with $\mathcal{C}\sim\mathcal{O}(1)$ in the condensation region. Recent multi-cluster comparisons further show that cool-core $\sigma_v$ values measured by XRISM tend to fall systematically below several cosmological simulation predictions \citep{AudardXRISM_2025}, qualitatively favouring gentler, less ejective feedback implementations that do not over-stir the atmosphere, in line with CCA-regulated self-regulation.

We also compare with \citet{gaspari2018}, who originally introduced both the turbulence-based $C$-ratio ($C\equiv t_{\rm cool}/t_{\rm eddy}$) and the k-plot as observationally testable diagnostics of CCA-driven multiphase condensation and fueling. Here we recover quantitatively consistent behaviour and extend these diagnostics to a fully resolved, time-dependent group-halo simulation. In our runs we find that the hot soft X-ray phase lies within it over the radii where filaments are expected to form and seed colder phases (e.g. $C\simeq0.5$--$2$ between $\sim5$ and $15$ kpc in \highc, and approaching the same band in \lowc\ as the halo evolves).  
Also, the \citet{gaspari2018} k-plot provides a clean observational mapping between ``ensemble-beam'' measurements, which average over many structures and therefore yield relatively large line broadening but small centroid shifts (typically $\sigma_{\rm los}\sim$ a few $\times10^{2}$ km s$^{-1}$ with $|v_{\rm los}|\lesssim$ several $\times10^{1}$ km s$^{-1}$), and ``pencil-beam'' measurements, which isolate individual clouds and can thus reveal narrow but fast components (mean $\sigma_{\rm los}\sim40$ km s$^{-1}$ with shifts reaching several $\times10^{2}$ km s$^{-1}$). Our macro-scale k-plot approaches the ensemble-beam locus, while exhibiting a slightly broader $\sigma_{\rm los}$ scatter during the stormy phase, as expected when extended multiphase coupling and stirring widen the range of turbulent velocities sampled along the line of sight. 
Our meso/micro-scale k-plots reproduce the pencil-beam phenomenology, we recover the same large spread towards high-$|v_{\rm los}|$ at modest $\sigma_{\rm los}$, corresponding to discrete infalling clouds and occasional HVC-like components.
The resulting progression from an ensemble-like locus on large scales to a more scattered, cloud-dominated locus on small scales provides a physical explanation for the two observational regimes highlighted by \citet{gaspari2018}, and shows explicitly how turbulence strength regulates the extent and overlap of the multiphase kinematic loci within the same CCA cascade.

Finally, independent constraints on AGN intermittency from ``echo'' phenomena provide a qualitative bridge between observed duty cycles and the CCA-driven variability characterised here. Ionisation echoes indicate that AGN activity can occur in discrete episodes with characteristic durations of order $\sim10^{5}$ yr \citep{Schawinski2015,keel2012}. This is compatible with a picture in which the nucleus is supplied by a sequence of stochastic multiphase accretion events rather than by a perfectly steady inflow. In this framework, the supply-side variability generated by condensation from meso to micro-scales can contribute to multi-timescale AGN flickering, while the shortest-timescale radiative variability is expected to arise within the unresolved accretion flow.

\section{Summary and conclusions}\label{conc}
In this work, we investigated SMBH feeding in turbulent, multiphase group-scale halos, focusing on how turbulence shapes the variability and kinematics of chaotic cold accretion (CCA) across the meso-scale. Building on the thermodynamic and morphological analysis presented in the companion {\sc BlackHoleWeather} study B26a, we examined here the radial inflow structure, the accretion history onto the SMBH, and the statistical properties of the resulting variability. Using high-resolution 3D hydrodynamical simulations with \athenapk, we followed the condensation cascade from tens of kpc scales down to the micro-scale region and quantified how different turbulence levels regulate both the spatial transport of multiphase gas and the temporal coherence of SMBH fueling. Our main results can be summarised as follows.
\begin{itemize}
\setlength{\itemsep}{0.3em}
\setlength{\parskip}{0pt}
\setlength{\parsep}{0pt}
    \item[(i)] In the absence of radiative cooling, turbulent accretion remains sub-Bondi and relatively steady, characteristic of hot-mode feeding. Once cooling is included, the system transitions to CCA and the accretion rate rises by more than an order of magnitude. Thanks to the sub-parsec resolution, we directly follow cold clumps and filaments down to $\sim 0.1$ pc and measure the SMBH feeding rate without relying on subgrid Bondi prescriptions. In both \highc\ and \lowc, the mean accretion rate remains strongly super-Bondi, with stochastic excursions spanning up to $\sim 2$ dex. Despite their different turbulence levels and filamentary morphologies, both fiducial simulations ultimately reach comparable mean SMBH accretion rates.

    \item[(ii)] Turbulence regulates the spatial extent and efficiency of cold-gas transport at kpc and meso-scales: strong stirring sustains extended, fragmented multiphase inflows, whereas weaker stirring yields smoother and more centrally concentrated transport. However, these order-of-magnitude differences in cold inflow at large radii do not translate into proportional differences at the smallest scales. Towards the centre, the inflow reorganises and converges to similar feeding efficiencies, indicating that SMBH accretion is governed primarily by meso-scale regulation rather than being directly supply-limited by the type of macro-scale weather.

    \item[(iii)] The radial inflow profiles reveal a scale-dependent structure of CCA. In the strongly stirred run (\highc), two distinct enhancements emerge: a meso-scale bump (0.1--1 kpc) associated with extended, fragmented multiphase transport, and a second inner bump at the innermost resolved scales tracing micro-scale accretion. In contrast, the weakly stirred case (\lowc) lacks the meso-scale amplification and shows only the inner accretion bump. This demonstrates that the stormy regime primarily boosts transport at intermediate radii, while both systems ultimately converge towards a common, centrally regulated feeding mode at small scales. Stormy and rainy weather are therefore not separate accretion modes, but scale-dependent expressions of the same condensation-driven cascade. 

    \item[(iv)] Although CCA drives strongly super-Bondi inflow, the Eddington-normalised accretion rate remains low for most of the time, with PDFs peaking at a few $\times 10^{-4}$ and extending through a broad tail towards higher $\lambda$. High-$\lambda$ episodes ($\lambda \gtrsim 10^{-2}$, e.g. quasar-like) occur only for $\sim 1\%$ of the time, confirming that the system predominantly resides in a low-accretion, maintenance-mode state (at least at low redshift). CCA therefore provides a natural mechanism to reconcile sustained multiphase fueling with predominantly low-Eddington black hole growth in radio-/maintenance-mode systems.

    \item[(v)] The power spectral density of the accretion history is well described by a broken power law, with pink-noise behaviour on long/intermediate timescales ($P(f)\propto f^{-1}$, as previously found in CCA studies; \citealt{gaspari2017}) and a steeper red-noise slope at high frequencies. Consistent with the fitted break frequencies ($f_{\rm b}\simeq 30~{\rm Myr}^{-1}$ for \highc\ and $f_{\rm b}\simeq 284~{\rm Myr}^{-1}$ for \lowc), the \lowc\ run preserves the $f^{-1}$ regime to higher frequencies, that is, to shorter timescales, than \highc. This indicates that weaker turbulence allows longer-lived, more coherent inflow fluctuations, whereas stronger turbulence accelerates decorrelation and damps short-timescale variability. The corresponding break times (\(\sim 3.5\)--33 kyr) are consistent with an inner parsec-scale damping/decorrelation transition, plausibly connecting halo-scale condensation to the temporal coherence of SMBH feeding and to the collisional, chaotic nature of CCA.

    \item[(vi)] CCA diagnostics provide a direct bridge between the thermodynamics of condensation and the observable kinematics of the multiphase halo. The multiphase $\mathcal{C}$-ratio profiles identify the soft X-ray phase as the key thermodynamic gateway of condensation: it is this phase that most persistently occupies the canonical $\mathcal{C}\sim 1$ band at radii of a few to several kpc, marking the region where nonlinear cooling is most efficient. By contrast, the cooler phases rapidly evolve towards $C\ll 1$, marking their transition into the lower-scale, top-down condensation cascade. In this regime they remain coupled within the multiphase rain, but trace progressively denser and more localised structures that channel gas towards the micro-scales.
    The multiphase k-plot provides a robust observational test of CCA and condensation, separating ensemble-like turbulent coupling from pencil-beam cloud infall, as originally proposed by \citet{gaspari2018} and corroborated here with fully resolved, time-dependent simulations. In our runs, the clearest stormy versus rainy separation emerges on meso-scales: \highc\ develops broad, overlapping multiphase loci with large velocity offsets and dispersions, whereas \lowc\ remains confined to a more compact and quiescent locus. Together, the $\mathcal{C}$-ratio and k-plot form a physically grounded, observation-ready framework to interpret the diversity of multiphase morphologies and kinematics in group and cluster cores and to discriminate between condensation-driven CCA fueling and quiescent hot-mode accretion.

\end{itemize}

Overall, together with B26a, this work builds on and extends the CCA framework developed in earlier theoretical and numerical studies \citep[e.g.][]{gaspari2013,gaspari2017,gaspari2018,GaspariTombesi2020} by providing a more unified, multiscale view of SMBH feeding in turbulent group-scale halos. By jointly analysing radial inflow rates, micro-scale accretion, variability statistics, and observation-ready kinematic diagnostics within the same simulations, we show that turbulence regulates not only the morphology of the multiphase medium, but also the temporal coherence and spatial transport of black-hole fueling. A key new result is that the meso-scale acts as the critical bridge between halo condensation and sub-parsec feeding: it is here that turbulence most strongly determines whether the atmosphere develops an extended, stormy multiphase cascade or a compact, rainy inflow, even though both ultimately converge towards locally regulated SMBH accretion at the smallest scales. In this sense, the present work confirms the core CCA picture while extending it towards a more explicit characterisation of the multiscale transport, variability, and kinematic signatures of feeding. Together, these results provide a quantitative baseline for future {\sc BlackHoleWeather} studies including additional physics, and for direct comparison with multi-wavelength observations of multiphase halo weather and AGN variability.\\

\begin{acknowledgements}
The BHW authors acknowledge key funding support from the European Research Council (ERC) under the European Union's Horizon Europe research and innovation programme (Consolidator Grant BlackHoleWeather, No.~101086804; PI: Gaspari). Views and opinions expressed are, however, those of the author(s) only and do not necessarily reflect those of the European Union or the European Research Council Executive Agency; neither the European Union nor the granting authority can be held responsible for them. We acknowledge ISCRA for awarding this project access to the LEONARDO supercomputer, owned by the EuroHPC Joint Undertaking, hosted by CINECA (Italy).
The numerical work was in part supported by the NASA High-End Computing (HEC) Program through the NASA Advanced Supercomputing (NAS) Division at Ames Research Center. 
VO acknowledges support from the DICYT ESO-Chile Comite Mixto PS 1757, and Fondecyt Regular 1251702. 
FMM acknowledges support from the Next Generation EU funds within the PNRR, Mission 4 - Education and Research, Component 2 - From Research to Business (M4C2), Investment Line 3.1 - Strengthening and creation of Research Infrastructures (project IR0000034 “STILES”).
MF acknowledges funding by the Deutsche Forschungsgemeinschaft (DFG, German Research Foundation) under Germany's Excellence Strategy -- EXC 2121 ``Quantum Universe'' --  390833306.
PT acknowledges support from NASA NNH22ZDA001N Astrophysics Data and Analysis Program under award 24-ADAP24- 0011.
RS acknowledges funding from the CAS-ANID grant No.~220016.
We thank Philipp Grete for support with the \athenapk\ code. 
We thank the organisers and participants of the following conferences for the stimulating discussions that helped improve this work: `BlackHoleWeather I' (Sexten, ITA), `Modelling of Multiphase Astrophysical Media' (Ringberg, GER), `Multi-phase, Multi-temperature, and Complex' (Olbia, ITA).

\end{acknowledgements}

   \bibliographystyle{aa}
   \bibliography{biblio.bib}

\end{document}